\documentclass{article}

\usepackage{jheppub}
\usepackage[english]{babel}

\usepackage{amsmath,amssymb}
\usepackage{graphicx}
\usepackage{dsfont,mathrsfs}

\DeclareMathOperator{\Tr}{Tr}
\DeclareMathOperator{\sgn}{sgn}
\DeclareMathOperator{\Ad}{Ad}
\DeclareMathOperator{\ad}{ad}
\DeclareMathOperator{\Sch}{Schw}
\DeclareMathOperator{\Pf}{Pf}
\newcommand{\Diff}{\operatorname{Diff}(S^1)}
\newcommand{\RR}{\mathds{R}}
\newcommand{\CC}{\mathds{C}}

\newcommand{\id}{\mathds{1}}
\newcommand{\pv}{\mathrm{p.v.}}
\newcommand{\orbit}{\mathscr{O}}
\newcommand{\scri}{\mathscr{I}}

\title{A menagerie of Schwarzians: coadjoint orbits of Virasoro and near-dS$_2$ quantum gravity}

\author{Henry Maxfield}
\affiliation{Leinweber Institute for Theoretical Physics at Stanford, 382 Via Pueblo, Stanford, CA 94305, USA}
\emailAdd{henrym@stanford.edu}

\abstract{
The Schwarzian theory, which governs the universal low-energy dynamics of near-extremal black holes and the SYK model, can be characterised as an integral over a particular coadjoint orbit of the Virasoro group. We describe and solve a complete classification of all possible generalised Schwarzian theories, defined by integrals over any Virasoro coadjoint orbit, including new classes of theories with qualitatively novel features. The classification of coadjoint orbits coincides with the moduli space of constant positive curvature two-dimensional Lorentzian geometries, and the associated Schwarzian theories govern associated wavefunctions in asymptotically near-dS$_2$ gravity (Jackiw-Teitelboim gravity in particular). The novel theories are inherently Lorentzian, defined by oscillatory path integrals weighted by $e^{iI}$ and force consideration of varying `coupling functions' (renormalised dilaton) which may not have definite sign. The definition of the theories involves an ambiguity, arising because the operator describing quadratic fluctuations at one loop fails to be essentially self-adjoint. This requires a choice of boundary condition, and also forces us to allow certain singularities in configurations and classical solutions. The choice is justified from  the realisation in JT gravity, which naturally regulates these singularities. The path integral remains one-loop exact via fermionic localisation, but this requires additional input beyond the Duistermaat-Heckman theorem.  This allows an exact computation of the path integral for all theories and all couplings, including new results for the original Schwarzian theory.
}

\begin{document}
\maketitle


\section{Introduction}

Many aspects of black holes at low temperature have been newly understood by the realisation that their dynamics has a universal sector governed by the Schwarzian theory, which also governs the  SKY model of disordered fermions at low energy \cite{Maldacena:2016hyu,Almheiri:2014cka,Jensen:2016pah,Engelsoy:2016xyb,Maldacena:2016upp,Sachdev:1992fk}. The Schwarzian theory describes a pseudo-Goldstone mode, arising when the infinite-dimensional symmetry group of diffeomorphisms of the circle $\Diff$ is broken to $PSL(2,\RR)$. For black holes this corresponds to breaking the group of reparameterisation of time to the symmetries of the AdS$_2$ spacetime which appears near the horizon of near-extremal black holes. This universal sector of near-AdS$_2$ is described by the negative-curvature version of the Jackiw-Teitelboim theory of two-dimensional dilaton gravity \cite{Jackiw:1984je,Teitelboim:1983ux}. The same Schwarzian also shows up in a different guise for the positive curvature JT gravity theory, where it governs the wavefunction for the Hartle-Hawking state of near-dS$_2$ spacetimes at late time \cite{Maldacena:2019cbz,Cotler:2019nbi}.

The Schwarzian theory has a nice mathematical characterisation as an integral over a particular coadjoint orbit of the Virasoro group \cite{Stanford:2017thb}. From this perspective, it is natural to ask about generalisations to different orbits. One class of generalisations has already proved important, where $\Diff$ is broken to a $U(1)$ subgroup: in AdS JT gravity these arise on hyperbolic trumpets \cite{Saad:2019lba}, or on hyperbolic geometries containing a conical defect \cite{Mertens:2019tcm}. But this does not exhaust the classification of coadjoint orbits \cite{Witten:1987ty,Balog:1997zz,lazutkin1975normal}: are there sensible theories corresponding to the remaining orbits, and do they have a physical realisation?

We answer both these questions in the affirmative, describing a complete family of Schwarzian theories in one-to-one correspondence with Virasoro coadjoint orbits,\footnote{The possibility of additional Schwarzian theories for the extra coadjoint orbits was previously pointed out in \cite{Mertens:2019tcm}.} and realising all of them in de Sitter JT gravity (as discovered in \cite{Held:2024rmg}). These orbits and Schwarzian theories are in one-to-one correspondence with the moduli space of constant positive curvature two-dimensional Lorentzian geometries (or more precisely, those containing a future asymptotic infinity $\scri_+$), which is the configuration space of dS JT \cite{Held:2024rmg,Alonso-Monsalve:2024oii,Alonso-Monsalve:2025lvt}. 

\begin{figure}
\centering
\includegraphics[width=.9\textwidth]{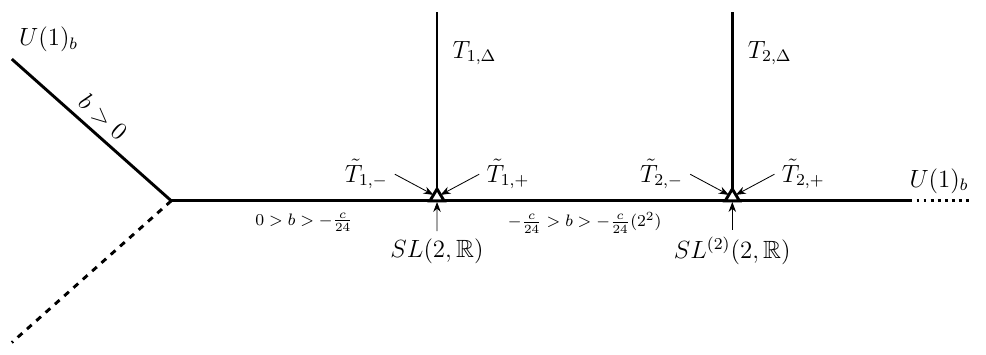}
\caption{The moduli space of Schwarzian theories, of Virasoro coadjoint orbits, and of constant positive curvature Lorentzian 2D geometries. The dashed line in the bottom left is a family of constant curvature geometries  (`big crunch' spacetimes), but does not represent a separate Schwarzian theory or coadjoint orbit since the spacetime does not contain an asymptotic future infinity $\scri_+$. The top left branch and horizontal spine consist of $U(1)_b$ orbits parametrised by a real constant $b$, with special $SL^{(n)}(2,\RR)$ points at $b=-\frac{c}{24}n^2$ for positive integers $n$ where the symmetry is enhanced. At each such vertex, a branch of `hyperbolic' $T_{n,\Delta}$ orbits (parametrised by $\Delta>0$) emerges. These vertices in fact split into three separate points (indicated schematically by the edges of the small triangles): a pair of exceptional `parabolic' orbits $\tilde{T}_{n,\pm}$ as well as the $SL^{(n)}(2,\RR)$ orbit.}\label{fig:moduli}
\end{figure}

This moduli space of Schwarzians or coadjoint orbits or geometries is shown in figure~\ref{fig:moduli}. This is locally one-dimensional (except at isolated special points). The previously studied Schwarzians lie on the central `spine' of $U(1)_b$ orbits leaving an unbroken $U(1)$, parametrised by a single real constant $b$. Among these orbits are special points labelled by positive integers $n$ where the unbroken symmetry is enhanced to $SL^{(n)}(2,\RR)$ (invariant under the $n$th Fourier mode); for $n=1$ this is the original $\Diff/PSL(2,\RR)$ theory. At each of these points emerges a new branch $T_{n,\Delta}$ labelled by $n$ and a positive real parameter $\Delta$, invariant under $\RR$ (a non-compact one-parameter subgroup of $\Diff$). Finally, each of these vertices in fact contains three distinct orbits: a pair of exceptional `parabolic' orbits $\tilde{T}_{n,\pm}$ in addition to the $SL^{(n)}(2,\RR)$ theories.

The remainder of this introduction summarises the main points of the paper: a complete classification of possible Schwarzian theories, the solution of the new theories, and their embedding in JT de Sitter quantum gravity. There are several qualitative ways in which the new theories differ from most previous work, and we emphasises these novelties. In particular, we will encounter technical subtleties which render the definition of the theory ambiguous. We primarily focus on one natural choice of completion of the theory, with a justification from JT gravity.

 \subsection{The Schwarzian theories and their novel features}

The most general Schwarzian theory is governed by an action depending on a diffeomorphism of the circle $\phi\in\Diff$:
\begin{equation}\label{eq:SchI}
    I_{b}[u,\phi] = \int \frac{d\theta}{2\pi} \, u(\theta) \left[\phi'(\theta)^2 b(\phi(\theta))-\frac{c}{12}\Sch(\phi,\theta)\right],
\end{equation}
where $\Sch$ denotes the Schwarzian derivative
\begin{equation}\label{eq:Sch}
    \Sch(\phi,\theta) = \left(\frac{\phi''(\theta)}{\phi'(\theta)}\right)' - \frac{1}{2}\left(\frac{\phi''(\theta)}{\phi'(\theta)}\right)^2,
\end{equation}
where primes denote differentiation with respect to $\theta$. The theory is characterised by  a function of the circle $b(\theta)$, and another function $u(\theta)$ which we think of as a source or coupling of the theory, and refer to as a `coupling function'.\footnote{There is no non-trivial dependence on the central charge $c$, since this can be absorbed into rescaling $b$ and $u$ as long as $c\neq 0$.}

Previously, both $b(\theta)$ and $u(\theta)$ have typically taken to both be constants: in particular, $u$ is proportional to the temperature for black holes or SYK. The basic reason is that the previously studied theories all have an unbroken $U(1)$ symmetry (perhaps part of a larger $PSL(2,\RR)$), and by choice of coordinates this can be taken to be the rigid rotations $\phi(\theta)=\theta+\theta_0$ (generated by $\partial_\theta$); invariance of the action then requires constant $b$. But the new theories we study have a qualitatively different symmetry group, which requires non-constant $b$. For related reasons, it is essential to allow the `coupling function' $u(\theta)$ to have non-trivial $\theta$-dependence. This has been considered before with the restriction that $u(\theta)>0$ \cite{Cotler:2023eza}, in which case a change of coordinate reduces to the constant $u$ case. But we will be forced to allow qualitatively different forms of $u(\theta)$ which oscillate between positive and negative values as a function of $\theta$. Considering such coupling functions will also give us new results for the original $\Diff/PSL(2,\RR)$ theory.

It turns out that the action $I_b[u,\phi]$ is bounded from below (as a function of the diffeomorphism $\phi$) only for those theories which can be described with constant $b\geq -\frac{c}{24}$, and only when $u(\theta)>0$ for all $\theta$. Restricting to this case allows only the previously-considered $\Diff/U(1)$ Schwarzians (with the extreme value $b=-\frac{c}{24}$ giving the original $\Diff/PSL(2,\RR)$ theory). This means that further generalisations do not make sense for a `Euclidean' theory defined by an integral weighted by $e^{-I}$ (relevant for computing thermal partition functions of AdS black holes, for example). Nonetheless, an unbounded action is not an obstruction to a `Lorentzian' theory weighted by $e^{iI}$, where we can potentially make sense of a conditionally convergent integral. With this in mind, the most basic quantity for our generalised Schwarzian theories is the `wavefunctional' $\Psi_b$ defined by
\begin{equation}\label{eq:SchwInt}
	\Psi_b[u] = \int \frac{\mathcal{D}\phi}{G_b} e^{i I_{b}[u;\phi]}.
\end{equation}
 The integral runs over $\Diff$, modulo a subgroup $G_b$ of diffeomorphisms which leave the action invariant (with a particular measure left implicit here, described below). This is precisely the sort of integral that shows up in de Sitter JT gravity, where it computes a wavefunction of the universe (as summarised in section \ref{ssec:dSintro}).

The last novel feature appears at the points $\theta_0$ where $u(\theta_0)=0$, which can now arise because we are not restricting $u$ to be positive. Since $u$ plays the role of an inverse coupling constant, it is perhaps not so surprising that something dramatic can happen at these points where the effective local value of $\hbar$ blows up. We will find that a consistent solution of these theories will lead us to consider certain non-smooth behaviour of diffeomorphisms $\phi$ at such points, and corresponding singularities in $b_\phi(\theta)$:
\begin{equation}\label{eq:singIntro}
    \phi(\theta)- \phi(\theta_0) \sim \sgn(\theta-\theta_0)|\theta-\theta_0|^p, \qquad b_\phi(\theta)\sim \frac{c}{24}\frac{p^2-1}{(\theta-\theta_0)^2}, \qquad (p>0).
\end{equation}
It will be necessary to include such singular configurations in the path integral, in particular as classical solutions, to make a sensible self-consistent theory.

Given these comments, our analysis will have to depart from the standard Schwarzians in several ways:
\begin{itemize}
	\item The function $b(\theta)$ characterising the theory must be taken to be non-constant.
	\item We must study the theory as a functional of a coupling function $u(\theta)$, which is not restricted to be constant, or even positive.
	\item The integral defining the theory is weighted by a phase $e^{i I}$, not $e^{-I}$.
    \item Singularities \eqref{eq:singIntro} must be allowed in the path integral and classical solutions.
\end{itemize}
These features make the analysis technically rather trickier, but the payoff will be some interesting and qualitatively different features present in the novel theories.

\subsection{Schwarzians and coadjoint orbits of Virasoro}

To explain the key properties of these theories it is first useful to describe this integral abstractly in terms of coadjoint orbits of the Virasoro group. All details of the necessary background on Virasoro coadjoint orbits are given in section \ref{sec:background}.

From the coadjoint orbit perspective, $b(\theta)$ (along with the central charge $c$) is an element of the coadjoint representation of the Virasoro group (the central extension of $\Diff$). Acting with a diffeomorphism $\phi^{-1}$ maps $b(\theta)$ to $b_\phi(\theta)$, which is the contents of the square brackets in \eqref{eq:SchI} (so $b$ transforms like a multiple of the stress tensor of a 2D CFT). As we vary $\phi$, $b_\phi(\theta)$ sweeps out an orbit $\orbit(b) = \{b_\phi:\phi\in\Diff\}$ of the coadjoint representation, with redundancy under diffs in $G_b$ which leave the seed $b$ invariant. Thus, the coset space $\Diff/G_b$ which we integrate over in \eqref{eq:SchwInt} is equivalent to the coadjoint orbit $\orbit(b)$ of Virasoro. In particular, choosing a different starting $b$ on the same orbit gives us an equivalent theory: genuinely different Schwarzians are classified not by the full data of $b$, but by the coadjoint orbit $\orbit$ of Virasoro to which it belongs. We will therefore often write  $\Psi_b$  as $\Psi_\orbit$, replacing the label $b$ with an indication of the orbit (labelling a point on the diagram in figure~\ref{fig:moduli}).

The coupling $u$ is then interpreted as an element of the dual adjoint representation, namely a vector field $u(\theta)\partial_\theta$ on $S_1$ generating diffeomorphisms.\footnote{Plus a multiple $\alpha$ of the central element, which simply shifts the action by a constant $\alpha c$.} The action $I_b[u;\phi]$ in \eqref{eq:SchI} is simply the natural pairing $\langle b_\phi,u\rangle = \int \frac{d\theta}{2\pi}u(\theta)b_\phi(\theta)$ between the adjoint $u$ and co-adjoint $b_\phi$. Similarly to how the theory depends only on the orbit $\orbit(b)$ of $b$, $\Psi_\orbit[u]$ depends non-trivially only on the orbit of $u$ under the adjoint action, which is the conjugacy class of the vector field under diffeomorphisms. `Non-trivially' here means that it transforms with an `anomaly' phase coming from the central extension of Virasoro \cite{Cotler:2023eza}:
\begin{equation}\label{eq:anomIntro}
    \Psi_\orbit\left[\frac{u\circ\phi}{\phi'}\right] = \exp\left(-i\frac{c}{12} \int \frac{d\theta}{2\pi}\frac{u(\phi(\theta))}{\phi'(\theta)} \Sch(\phi,\theta)\right)\Psi_\orbit\left[u\right].
\end{equation}
If we restrict to vector fields with no zeroes ($u>0$ everywhere, say) then space of conjugacy classes is one-dimensional, parameterised by the diff-invariant integral $\int \frac{d\theta}{u}$, and every conjugacy class has a representative with constant $u$: this explains why it is sufficient to consider only constant coupling if we require $u>0$. But if we relax this restriction we get new families of conjugacy classes where $u$ has $2n$ simple zeroes (in the generic case without higher-order zeroes);  the discrete label $n$ is constant on a conjugacy class, and the invariant $\int \frac{d\theta}{u}$ also generalises to this case when defined using the Cauchy principle value prescription.\footnote{Under smooth diffeomorphisms, there are $2n$ additional real invariants given by the values $u'(\theta_k)$ at the zeroes $u(\theta_k)=0$, up to cyclic permutation. But the non-smooth diffeomorphisms mentioned above do not preserve $u'(\theta_k)$. After allowing such diffs, the space of conjugacy classes with $2n$ simple zeroes becomes one-dimensional, parametrised only by $\int \frac{d\theta}{u}$.} The upshot is that the Schwarzian theories can (and do) give qualitatively different results in sectors where $n$ takes different values: in fact, it turns out that the path integral \eqref{eq:SchwInt} will be non-zero only in one such sector (or two sectors for the special $SL^{(n)}(2,\RR)$ orbits including the original $\Diff/PSL(2,\RR)$ theory).



Coadjoint orbits naturally carry the structure of a symplectic manifold (which provides the measure used in \eqref{eq:SchwInt}), and this allows us to interpret the action $I_b[u,\phi]$ as the Hamiltonian function for the flow generated by the action of $u$. This was the key observation of \cite{Stanford:2017thb}, which immediately allows the use of the Duistermaat-Heckman (DH) theorem \cite{Duistermaat:1982vw} to show that the integral \eqref{eq:SchwInt} is one-loop exact. However, an important requirement of this theorem is that the Hamiltonian generates a $U(1)$ flow (i.e., all orbits must be periodic with fixed finite period). This holds for $u>0$, since the vector field $u(\theta)\partial_\theta$ generates rigid rotations in coordinates $\tilde{\theta}$ with $d\tilde{\theta} \propto \frac{d\theta}{u(\theta)}$. But it fails when $u$ has zeroes: the flow is a non-compact $\RR$, so we cannot immediately apply the DH theorem. Nonetheless, we can patch the gap left by the $U(1)$ in the proof of DH by fermionic localisation (described in section \ref{sec:localisation}). The $U(1)$ symmetry is used in the proof to show existence of a symmetric bilinear form which is invariant under the flow (by averaging an arbitrary metric over the symmetry group). We bypass this step by instead simply explicitly exhibiting such a form (constructed from the $u$-invariant pairing of vector fields $\Gamma(v_1,v_2) = \int \frac{v_1 v_2}{u^3}$, defined with a principle value prescription), which we use to construct a deformation which localises the path integral onto classical solutions. After addressing technical subtleties with this argument (related to the class of singularities \eqref{eq:singIntro} we allow), the conclusion is that the path integral \eqref{eq:SchwInt} nonetheless remains one-loop exact.

\subsection{Solving the new Schwarzians}

Given the above argument that the path integral \eqref{eq:SchwInt} is one-loop exact, to evaluate the Schwarzian integrals it is sufficient to determine all possible classical solutions and to compute the determinants governing quadratic fluctuations around those saddle-points. This solutions is described in section \ref{sec:solution}. We will focus here on the orbits $T_{n,\Delta}$ to illustrate the main new features.

The classical solutions are stationary points of the action $\frac{\delta}{\delta\phi}I_b[u;\phi]=0$. In the description as a coadjoint orbit integral, the classical solutions are points $b_\phi$ on the orbit $\mathcal{O}(b)$ which are left invariant by the action of the `coupling' $u$ (interpreted as a Lie algebra element). So, a classical solution $b(\theta)$ is a function which is invariant under the coadjoint action of the infinitesimal diffeomorphism $u(\theta)$. Given $u$, it is straightforward to explicitly solve for such $b(\theta)$, and we subsequently impose additional constraints to ensure that $b$ lies in the desired coadjoint orbit $\orbit$. In particular, in the case $\orbit=T_{n,\Delta}$ we find the following:
\begin{equation}\label{eq:ClassSolIntro}
    \text{Classical solution in $T_{n,\Delta}$}:\qquad b(\theta) = \frac{c}{24}\frac{2u(\theta) u''(\theta)-(u'(\theta))^2+\lambda^2}{u(\theta)^2}, \qquad \Delta = \lambda \int_\pv\frac{d\theta}{u(\theta)},
\end{equation}
where $u$ is required to have exactly $2n$ simple zeroes (and there are no classical solutions in other cases). But this classical solution is singular unless we impose additional constraints on $u$: specifically, we require that $u'(\theta) = \pm \lambda$ at points where $u(\theta)=0$. This means that the $T_{n,\Delta}$ theories only have smooth solutions on a submanifold in the space of couplings of codimension $2n$. At this stage it is unclear whether we should strictly constrain ourselves to these smooth solutions, or whether to allow singularities.

The next step is to consider the one-loop fluctuations around a smooth solution (choosing to evaluate $\Psi_{T_{n,\Delta}}[u]$ at a coupling $u$ which satisfies the $2n$ constraints on $u'$ at its zeroes). This can be expressed in terms of the determinant $\det Q$ of an operator $Q$ acting on a Hilbert space $L^2(S^1)$ of functions on the circle (along with a similar determinant accounting for the symplectic measure). The functions in question are the infinitesimal diffeomorphisms $v(\theta)$ which generate $\Diff$, modulo multiples of $u(\theta)$ (which is the generator of the stabiliser $G_b$). Now we find a surprise: the quadratic fluctuation (Lichnerowicz) operator $Q$ fails to be essentially self-adjoint! This is due to singular behaviour at the zeroes of $u$ (specifically, vanishing of the coefficient of the highest derivative term in $Qv(\theta) = u(\theta)v''''(\theta)+\cdots$). In down-to-earth terms, to unambiguously define $Q$ as a Hermitian operator (and in particular to make sense of $\det Q$), we are forced to make a choice of boundary conditions at the points $\theta_0$ where $u(\theta_0)=0$.  Any choice of boundary condition will inevitably allow for certain singularities in $v(\theta)$ (and disallow others). There are several possible natural choices one could make; we select one of these guided by dS$_2$ JT gravity, which allows singularities of the form $v(\theta) \sim (\theta-\theta_0)\log |\theta-\theta_0|$. With this choice, we can evaluate the one-loop path integral around the smooth solution. We comment on alternative choices of completion in sections \ref{ssec:altbcs} and \ref{ssec:discbcs}.

This choice of boundary conditions for the one-loop fluctuations has consequences for the space of classical solutions \eqref{eq:ClassSolIntro}. Note that the singular $v(\theta)$ mentioned above is precisely the infinitesimal generator of the singular diffeomorphism given in \eqref{eq:singIntro} ($v$ is the derivative of $\phi$ with respect to $p$ evaluated at $p=1$). By considering generic perturbations $\delta u$ of the coupling around the smooth solution, we find that our boundary conditions allow us to move off the manifold $b_\phi \in \orbit$ spanned by smooth $\phi\in\Diff$, turning on singularities of the form $b\sim (\theta-\theta_0)^{-2}$. Hence, the singular classical solutions in \eqref{eq:ClassSolIntro} become admissible for any $u$!


These considerations generalise to all other orbits and all possible couplings $u$. First, we find that the path integral for the parabolic $\tilde{T}_{n,\pm}$ orbits is most naturally defined as a limit of the hyperbolic $T_{n,\Delta}$ orbits as $\Delta\to 0$, with the two theories differing by a restriction on the sign of $\int\frac{1}{u}$. For the $SL^{(n)}(2,\RR)$ case, we note that the classical solutions \eqref{eq:ClassSolIntro} (singular or otherwise) can never belong to these orbits if $\int\frac{1}{u}\neq 0$, so the path integral must vanish. However, \eqref{eq:ClassSolIntro} always gives a representative of this orbit when the constraint $\int\frac{1}{u}=0$ is satisfied (and $u$ has $2n$ simple zeroes), for any $\lambda$. So, there is a moduli space of classical solutions parametrised by $\lambda$ (generated by a zero mode of the quadratic fluctuation operator $Q$), and integrating over this moduli space gives us a delta function $\delta(\int\frac{1}{u})$ in the path integral. Finally, the other $U(1)_b$ orbits never have classical solutions when $u$ has zeroes, and the corresponding path integrals must vanish.

\begin{table}[t]
\centering
\renewcommand{\arraystretch}{2.5}
\begin{tabular}{c c c}
\hline\hline
Orbit $\orbit$ & Coupling $u(\theta)$ & $\Psi_\orbit[u]$ \\
\hline
$U(1)_b$ & $u>0$ &
$\displaystyle \left(\int\!\frac{d\theta}{u}\right)^{\!-1/2} \exp\left[2\pi i b \left(\int\!\frac{d\theta}{u}\right)^{\!-1} \!- i\frac{c}{24}\int\!\frac{d\theta}{2\pi}\frac{(u')^2}{u}\right]$ \\[6pt]
$SL^{(n)}(2,\RR)$ & $u>0$ &
$\displaystyle \left(\int\!\frac{d\theta}{u}\right)^{\!-3/2} \exp\left[-2\pi i \frac{c}{24}n^2 \left(\int\!\frac{d\theta}{u}\right)^{\!-1} \!- i\frac{c}{24}\int\!\frac{d\theta}{2\pi}\frac{(u')^2}{u}\right]$ \\[6pt]
$T_{n,\Delta}$ & $2n$ simple zeroes &
$\displaystyle \left(\int\!\frac{d\theta}{u}\right)^{\!-1/2} \exp\left[i\frac{c}{24}\!\left(\frac{\Delta^2}{2\pi}\left(\int\!\frac{d\theta}{u}\right)^{\!-1} \!- \int\!\frac{d\theta}{2\pi}\frac{(u')^2}{u}\right)\right]$ \\[6pt]
$\tilde{T}_{n,\pm}$ & $2n$ simple zeroes &
$\displaystyle \left(\int\!\frac{d\theta}{u}\right)^{\!-1/2} \exp\left[-i\frac{c}{24}\int\!\frac{d\theta}{2\pi}\frac{(u')^2}{u}\right] \Theta\!\left(\pm\int\!\frac{1}{u}\right)$ \\[6pt]
$SL^{(n)}(2,\RR)$ & $2n$ simple zeroes &
$\displaystyle \delta\!\left(\int\!\frac{d\theta}{u}\right) \exp\left[-i\frac{c}{24}\int\!\frac{d\theta}{2\pi}\frac{(u')^2}{u}\right]$ \\[6pt]
\hline\hline
\end{tabular}
\caption{Results for the Schwarzian path integral $\Psi_\orbit[u]$ for all coadjoint orbits $\orbit$ and coupling functions $u(\theta)$. The first two rows are previously known results (and also apply for $u<0$. In all other cases (in particular, for $T_{n,\Delta}$ or $\tilde{T}_{n,\pm}$ with $u>0$, for generic $U(1)_b$ with $u$ having zeroes, or when $u$ has $2m\neq 2n$ simple zeroes), the path integral vanishes: $\Psi_\orbit[u]=0$. Where $u$ has zeroes, all integrals are defined with a principal value prescription.}\label{tab:results}
\end{table}

We summarise the results for the path integrals $\Psi_\orbit[u]$ for all possible orbits $\orbit$ and for all possible coupling functions $u$ in table \ref{tab:results}.

\subsection{de Sitter JT gravity}\label{ssec:dSintro}

The full classification of Schwarzian theories described here is realised in the 2D gravity model of de Sitter JT. This theory has a finite-dimensional configuration space matching the diagram in figure~\ref{fig:moduli}, each point corresponding to a different constant positive curvature Lorentzian 2D metric $g$.

The Schwarzian theory emerges when we compute solutions to the Wheeler-DeWitt equation corresponding to a fixed geometry $g$ in a limit approaching future infinity $\scri_+$. This can be described as a path integral with boundary conditions giving the intrinsic data of a Cauchy surface: its length $2\pi a$ and the dilaton profile $\Phi(\theta)$ (where $a\theta$ is proper length). In a limit where we take $a\to\infty$ and fix the `renormalised dilaton' $u(\theta) = a^{-1}\Phi(\theta)$, this is computed by a Schwarzian theory:
\begin{equation}\label{eq:PsiWDW}
    \Psi^\text{WDW}_g[a,\Phi] \sim \exp\left[i\int\frac{d\theta}{2\pi}a\Phi\right] \Psi^\text{Schw}_\orbit[u = \tfrac{\Phi}{a}],\qquad a\to \infty, \; u \text{ fixed},
\end{equation}
where the orbit $\orbit$ in the Schwarzian theory is selected by the locally dS$_2$ geometry $g$. This was demonstrated for the specific case $g=dS_2$ of de Sitter space and $\orbit=SL(2,\RR)$ in \cite{Maldacena:2019cbz}, and for all constant $b$ orbits in \cite{Cotler:2019nbi}. We demonstrate this correspondence explicitly for the $T_{n,\Delta}$ orbits in section \ref{sec:JT}, and other orbits follow straightforwardly using the same ideas, at the level of the formally-defined path integral. The Wheeler-DeWitt equation for JT gravity reduces in this limit to the infinitesimal version of the transformation \eqref{eq:anomIntro} (analogously to the results of \cite{Chakraborty:2023yed} for Einstein gravity in higher dimensions).

We emphasised above that the formal path integral is not enough to give a unique definition of the Schwarzian theory when $u(\theta)$ has zeroes; we made a choice of boundary condition which in particular allows the singular configurations in \eqref{eq:singIntro}. One justification is that these singularities get regularised in JT gravity when the cutoff $a$ (taken to infinity in the Schwarzian limit) is restored, as we show in section \ref{sec:JTnonsmooth}. At large but  finite $a$ the singularities \eqref{eq:singIntro} become milder, with finite and unambiguous action.

\section{Background: coadjoint orbits of Virasoro}\label{sec:background}

\subsection{Coadjoint orbits}

Consider first a general Lie group $G$ with Lie algebra $\mathfrak{g}$, writing its elements as $u,v,\ldots$ and elements of the dual space as $\mathfrak{g}^*$ as $a,b,\ldots$. We write the pairing between $\mathfrak{g}^*$ and $\mathfrak{g}$ as $\langle a,u\rangle\in \RR$. The adjoint and coadjoint representations of the Lie algebra $\mathfrak{g}$ are defined by 
\begin{equation}
    \ad_u(v) = [u,v],\qquad \langle \ad^*_u(a), v \rangle = -\langle a, [u,v]\rangle.
\end{equation}
The latter equation just says that the coadjoint $\ad^*_u$ is  minus the transpose of the adjoint $\ad_u$.
These exponentiate to representations $\Ad$, $\Ad^*$ of the group $G$,
\begin{equation}
    \Ad_g(v) = v_g = g v g^{-1},\qquad   \Ad^*_g(a) = a_g, \quad \langle a_g, v\rangle = \langle a, \Ad_{g^{-1}}(v)\rangle \,,
\end{equation}
so the coadjoint group representation is the inverse transpose of the adjoint.

The \emph{coadjoint orbit} $W_b$ containing $b$ is simply the orbit of $b$ under the coadjoint action: 
\begin{equation}
    W_b = \{\Ad^*_g(b):g\in G\}.
\end{equation}
The notation here is redundant; we could choose $b$ to be any representative of the orbit. The orbit is a homogeneous space $G/G_b$, where $G_b$ is the stabiliser of $b$ under the coadjoint action (the subgroup of $g$ such that $\Ad^*_g(b) = b$). Infinitesimally, this means that the tangent space of $W$ at $b\in W$ is identified with a quotient of the Lie algebra $\mathfrak{g}$ by the elements that leave $b$ invariant under the coadjoint action:
\begin{equation}
    T_b W \simeq \mathfrak{g}/ \{u: \ad^*_u b=0\}.
\end{equation}

Coadjoint orbits $\orbit$ are of particular interest because they are symplectic manifolds. Given the above identification of $T_b\orbit$, we can express the symplectic form in terms of Lie algebra elements:
\begin{equation}
    \omega_b(u,v) = \langle b, [u,v] \rangle.
\end{equation}
This is well-defined and non-degenerate on $T_b\orbit$ because $\iota_u\omega_b = \omega(u,\cdot) =  -\ad^*_u(b)$, which is zero if and only if $u$ corresponds to the zero vector in $T_b\orbit$. It is also closed, $d\omega=0$ (a simple exercise in chasing definitions, using the Jacobi identity).

Each Lie algebra element $u$ defines a function $I_{W,u}$ on the coadjoint orbit $W$, simply by the pairing $I_{W,u}(b) = \langle b, u\rangle$. Thinking of this function as a Hamiltonian on the symplectic manifold, it generates the flow of the coadjoint action of $u$ (that is, $dI_{W,u} = \iota_u \omega$).

\subsection{Coadjoint orbit `theories'}\label{ssec:coadjtheories}

Given a coadjoint orbit $\orbit$ and a Lie algebra element $u\in\mathfrak{g}$, we can define a `theory' whose action is the simple function defined above by pairing $u$ with coadjoint vectors:
\begin{equation}\label{eq:Iabstract}
    I_{\orbit,u}[b] = \langle b, u\rangle, \qquad b\in \orbit.
\end{equation}
 By `theory' here we just mean an integral of $ e^{-I_{\orbit,u}}$ or  $e^{iI_{\orbit,u}}$ over the orbit, or more generally `expectation values' obtained by inserting a function on $\orbit$. The measure in this integral is the symplectic measure $d\omega\wedge\cdots\wedge  d\omega$ (given by the Pfaffian of $\omega$). 

Equivalently, we can choose a single representative coadjoint $b\in\mathfrak{g}^*$ on the orbit $\orbit(b)$, and identify the orbit with the homogeneous space $G/G_b$, represented by cosets of elements of the group:
\begin{equation}
    I_{b,u}[g] = \langle \Ad^*_g b, u \rangle = \langle b, \Ad_{g^{-1}}(u)\rangle, \qquad g\in G/G_b.
\end{equation}

The classical solutions of such a theory are the stationary points of $I_{\orbit,u}[b]$ with respect to variations of $b$ within the orbit $\orbit$. A variation can be given by $\delta b = \ad^*_v b$ for some (non-unique) Lie algebra element $v$, and the corresponding variation of the action is $\delta I_{\orbit,u} = \omega_b(u,v)$, which vanishes for all $v$ when $\ad^*_u b =0$. So, classical solutions correspond precisely to fixed points of $u$ in the orbit (when $u$ is tangent to the stabiliser group $G_b$).



\subsection{The Virasoro group}

The Virasoro group is a central extension of the group $\Diff$ of diffeomorphisms of the circle. Diffeomorphisms are generated by vector fields $v = v(\theta)\partial_\theta$ (where $\theta\sim \theta + 2\pi$ us a coordinate on $S^1$. So, an element of the adjoint representation (i.e., the Virasoro algebra) is a pair $(v,\alpha)$ where is a vector field and $\alpha$ is a multiple of the central element $C$. The adjoint action of the Virasoro group by a diffeomorphism $\phi\in \Diff$ (being independent of the central part) is given by
\begin{equation}\label{eq:Adv}
    \Ad_{\phi^{-1}}(v(\theta),\alpha) = \left(\frac{v(\phi(\theta))}{\phi'(\theta)}, \alpha+\frac{1}{12} \int \frac{d\theta}{2\pi}\frac{v(\phi(\theta))}{\phi'(\theta)} \Sch(\phi,\theta) \right),
\end{equation}
where $\Sch$ is the Schwarzian derivative given in equation \eqref{eq:Sch}.
The infinitesimal version  of this (i.e., the adjoint representation of the Virasoro algebra) is
\begin{equation}\label{eq:adv}
    [u,v] = \left(u'(\theta)v(\theta)-u(\theta)v'(\theta), -\frac{1}{12}\int \frac{d\theta}{2\pi} u'''(\theta)v(\theta)\right).
\end{equation}
An element in the dual coadjoint representation is a pair $(b(\theta),c)$, which pairs with adjoint vectors as 
\begin{equation}
    \langle (b, c), (v,\alpha) \rangle = \int \frac{d\theta}{2\pi}b(\theta)v(\theta)+\alpha c \,. 
\end{equation}
The coadjoint action leaves $c$ invariant, so we can regard this as a fixed constant. The action on $b$ is
\begin{equation}\label{eq:coAdjaction}
    b_\phi(\theta) = \Ad^*_{\phi^{-1}}(b(\theta))=\phi'(\theta)^2 b(\phi(\theta))-\frac{c}{12}\Sch(\phi,\theta),
\end{equation}
familiar as the transformation of a stress tensor under conformal transformations (up to a factor of $-2$). The infinitesimal version (coadjoint representation of the Virasoro algebra) is
\begin{equation}\label{eq:coadjaction}
    \ad_u^*(b,c) = \left(-2u' b - u b'+ \frac{c}{12}u''',0\right).
\end{equation}

\subsection{Classification of Virasoro coadjoint orbits I: stabilizer}\label{ssec:classify}

We next review the necessary background on the classification of Virasoro coadjoint orbits (disregarding the special case $c = 0$). We will find it useful to have two different perspectives on this classification. The first proceeds (mostly following \cite{Witten:1987ty}) by identifying the stabiliser of a point on the orbit, namely Lie algebra elements (i.e., vector fields $u(\theta)$) which act trivially. The second method of classification (by the solution set of the Hill equation) will be presented in the next subsection.

So, we first aim to identify the vector fields $u$ which leave an element $b$ of the coadjoint representation invariant, so $\ad_u^*b=0$. The conjugacy class of such $u$ under diffeomorphisms is invariant over a coadjoint orbit, so this relates the classification of orbits to the classification of conjugacy classes in $\Diff$. From \eqref{eq:coadjaction}, we are looking for solutions to the differential equation
\begin{equation}\label{eq:invariance}
    \frac{c}{12}u''' -2u' b - u b'=0.
\end{equation}
 A topological argument \cite{Witten:1987ty} says that (for $c\neq 0$) the number of solutions modulo 2 is invariant under deformations of $b$. This implies (by deforming to $b=0$, for example) that there is always either a one- or three-dimensional space of solutions. In particular, for every $b$ there exists some non-zero $u$ solving \eqref{eq:invariance}, so a classification by looking for stabilising $u$ will not miss any orbits.


\subsubsection{The $U(1)_b$ orbits}

First we treat the case that where a point on our coadjoint orbit is invariant under some $u(\theta)$ with no zeroes, so $u$ is strictly positive (or strictly negative) everywhere.

Then, $u$ is conjugate to a constant vector field (by a diffeomorphism to coordinate $\tilde{\theta}$ with $d\tilde{\theta} \propto \frac{d\theta}{u(\theta)}$),  with the constant specified by the invariant integral $\int\frac{1}{u}$. So, the conjugacy class of a nowhere-zero vector field is determined completely by the value of the invariant $\lambda$. We also conclude that there is a point $b$ on the orbit which is stabilized by the constant vector field, and \eqref{eq:invariance} immediately tells us that $b$ is constant. So, every coadjoint which is invariant under an everywhere-positive vector field is in an orbit with a constant-$b$ representative.

The stabiliser group generated by such a $u$ is $G_b\simeq U(1)$ (conjugate to the rigid rotations $\theta \mapsto \theta + \theta_0$ for constant $\theta_0$). But this may be part of  a larger  stabiliser group. To check this, it is sufficient to solve \eqref{eq:invariance} for constant $b$. There are generically no additional $2\pi$ periodic solutions, with the exceptions occurring when $b=-\frac{c}{24}n^2$ for positive integers $n$. In these cases, the extra solutions $u(\theta)=\cos(n\theta)$ and $u(\theta) = \sin(n\theta)$ extend the stabiliser group to $G_b = PSL^{(n)}(2,\RR)$ (spanned by $L_{\pm, n}$ and  $L_0$ shifted by a central term, using the usual notation for the generators of Virasoro).

These orbits exhaust all the usual Schwarzian theories (making up the `spine' in figure~\ref{fig:moduli}), where the stabilizer $G_b$ is $U(1)$ or $PSL(2,\RR)$ (which contains a $U(1)$). These orbits are labelled by (constant) $b$  (along with $c$, which we leave implicit), so we name them $U(1)_b$. In particular, the case $b = -\frac{c}{24}$ is the original $PSL(2,\RR)$ Schwarzian theory.

\subsubsection{The `hyperbolic' $T_{n,\Delta}$ orbits}

The other classes of orbits appear when we consider $u$ which vanishes somewhere. Now there are more invariants characterising the conjugacy class of $u$. First, the number of zeroes is a discrete invariant. Second, smooth diffeomorphisms also leave invariant the derivative $u'$ at each zero, up to cyclic permutations of the zeroes (though we will later be compelled to consider non-smooth diffs which can change these values). Finally, assuming all zeroes are simple ($u'\neq 0$), the integral of $u^{-1}$ encountered above is still an invariant when defined with a principal value prescription (e.g., defined by the real part of $\int \frac{1}{u-i\epsilon}$). This exhausts the invariants, so a pair of vector fields with $2n$ simple zeroes are related by a smooth change of coordinates if and only if they have the same values of the $2n+1$ invariants. So, among vector fields with $2n$ simple zeroes there is a $(2n+1)$-dimensional space of conjugacy classes. We will return to cases with higher-order zeroes below. 

To see what coadjoints are invariant under a given $u$, we can regard \eqref{eq:invariance} as a differential equation for $b$. This has the general solution
\begin{equation}\label{eq:bfromu}
    b = \frac{c}{24}\frac{2u u''-(u')^2+\lambda^2}{u^2}
\end{equation}
with an arbitrary integration constant $\lambda^2$. Now, if $u$ has zeroes then the vanishing denominator is in danger of making $b$ blow up. Demanding smooth $b$ and avoiding these singularities will thus constrain the possibilities for both $u$ and $\lambda$ (though we will later find it necessary to relax these constraints).

Consider first a case where $u$ has at least one simple zero (with $u'\neq 0$). Then finiteness of $b$ in \eqref{eq:bfromu} at such a zero requires the numerator to vanish so $\lambda=|u'|$. This means that $|u'|$ must take the same value at every zero (and in particular there can only be simple zeroes). There must be an even number of zeroes $2n$, with $u'$ alternating between $\pm \lambda$.\footnote{We could worry about the possibility of nasty smooth functions with infinitely many zeroes, but this is impossible with $|u'|$ equal at every such point because $u'$ would fail to be continuous at an accumulation point of zeroes.} Among such vector fields with fixed $\lambda$, there remains only a one-dimensional space of conjugacy classes of $u$ up to overall normalisation, characterised by the invariant $\Delta$:
\begin{equation}\label{eq:Delta}
    \Delta = \lambda \int_{\mathrm{p.v.}} \frac{d\theta}{u(\theta)}.
\end{equation}
This quantity parametrises a family of coadjoint orbits $T_{n,\Delta}$ for positive integers $n$, and $\Delta> 0$: we can take $\Delta$ to be non-negative by flipping the sign of $u$ if necessary, and we will see in a moment that $\Delta=0$ returns us to previously encountered $SL^{(n)}(2,\RR)$ orbits.

For $\Delta>0$,  $u$ is the unique vector field which leaves $b$ invariant (up to rescaling). To see this, note that if \eqref{eq:bfromu} holds, then the other two solutions to \eqref{eq:invariance} are $u(\theta) \exp\left(\pm \lambda\int^\theta\frac{1}{u}\right)$ (using a principle value prescription at the zeroes of $u$). Under monodromy around the circle, these additional solutions pick up factors of $e^{\pm\Delta}$, so there is no extra periodic solution unless $\Delta=0$.

In the special case $\Delta=0$, we can represent the conjugacy class by $u(\theta) = \sin(n\theta)$. This simply gives us back the constant $b=-\frac{c}{24}n^2$ orbit we saw already, with $u$ being an extra generator of $SL^{(n)}(2,\RR)$. This means that the $U(1)_b$ branch of orbits and the $T_{n,\Delta}$ branch meet at $b=-\frac{c}{24}n^2$ and $\Delta=0$, as in figure~\ref{fig:moduli}. The different branches can be thought of as deformations of the  $SL^{(n)}(2,\RR)$ orbit, where the three-dimensional invariant subgroup is broken to a one-dimensional subgroup: if that subgroup is generated by an elliptic element we move onto the $U(1)_b$ branch, and if it is hyperbolic we move onto the $T_{n,\Delta}$ branch (the marginal case of parabolic elements will be addressed next). This precisely matches the ways that dS$_2$ (or its $n$-fold covers) can be deformed to different families of constant-curvature spacetimes by breaking the isometry group $\mathfrak{so}(1,2)$ to different one-parameter subgroups.


\subsubsection{The `parabolic' $\tilde{T}_{n,\pm}$ orbits}

Finally, we can consider the possibility that $u$ has a double zero. For smooth $b$, this means the integration constant in \eqref{eq:bfromu} vanishes, $\lambda=0$, and all zeroes are double zeroes (they cannot be higher order, because then  \eqref{eq:invariance} will imply that $u$ is identically zero). In particular, this means that $u$ cannot change sign (so we can choose $u\geq 0$ WLOG). Additionally, $u'''=0$ at all zeroes to avoid a single pole in $b$. The conjugacy classes of $u$ (up to rescaling) are labelled by the number of zeroes $n$ and a discrete parameter $\varepsilon\in\{+,-,0\}$. This parameter is once again related to a principal value integral of the form $\int\frac{1}{u}$:  in this case, the integral can be defined by integrating $\frac{1}{u\pm i\epsilon^2}$ and subtracting a singular part $\propto \epsilon^{-1}$ (where it is important that $u'''= 0$ so there is no logarithmic divergence). Then $\varepsilon$ is the sign of this integral (having made the choice $u\geq 0$).

The  $\varepsilon=0$ orbit is represented by $u(\theta)=1-\cos(n\theta)$, which once again gives us back the $b=-\frac{c}{24}n^2$ coadjoint orbit invariant under $PSL^{(n)}(2,\RR)$. So the conjugacy classes which give us new orbits are labelled by a sign, giving our last two classes of coadjoint orbits $\tilde{T}_{n,\pm}$.

\subsection{Classification of orbits II: Hill equation}\label{ssec:classifyHill}

We now review a second method to characterise the Virasoro coadjoint orbits \cite{lazutkin1975normal,Balog:1997zz}. This will be useful later, because we will be driven to somewhat relax our smoothness conditions on diffeomorphisms $\phi(\theta)$ and correspondingly allow certain singularities in our coadjoints $b(\theta)$, and this second method is more robust under this change.

This method uses a relation between the Virasoro coadjoint action and the Hill equation,
\begin{equation}\label{eq:Hill}
    \psi''(\theta) - \frac{6}{c}b(\theta) \psi(\theta) = 0\,.
\end{equation}
The key observation is that the local solutions to this equation (i.e., solutions without demanding periodicity, or solutions on the universal cover $\RR$ of $S^1$) are invariant under a coadjoint action on $b$ (as defined in \eqref{eq:coAdjaction}), along with a simultaneous action on $\psi$ as a field of weight $-\frac{1}{2}$:
\begin{equation}\label{eq:Hilltransform}
    \psi''(\theta) - \frac{6}{c}b(\theta) \psi(\theta) = 0 \iff \psi_\phi''(\theta) - \frac{6}{c}b_\phi(\theta) \psi_\phi(\theta)=0, \qquad \psi_\phi(\theta) = \frac{\psi(\phi(\theta))}{\sqrt{\phi'(\theta)}}.
\end{equation}

In particular, the monodromy of the Hill equation is the same for all $b$ in a given coadjoint orbit, in a sense we now explain. Let $\psi_{\pm}$ be a pair of linearly independent real solutions to \eqref{eq:Hill} and arrange them in a vector $\Psi(\theta) = (\psi_+(\theta),\psi_-(\theta))^t$. Then we define the monodromy matrix $M_\Psi$ by
\begin{equation}
    \Psi(\theta+2\pi) = M_\Psi \Psi(\theta), \qquad M_\Psi \in SL(2,\RR).
\end{equation}
To show that $M_\Psi$ is in $SL(2,\RR)$ (i.e., has unit determinant), we use the fact that the Wronskian $W = \det(\Psi(\theta)\;\Psi'(\theta))$ is independent of $\theta$. Now $M_\Psi$ itself does not characterise the equation because we have the freedom to choose a different basis of solutions, but its conjugacy class in $SL(2,\RR)$ is invariant, and depends only on the coadjoint orbit of $b$. So, the conjugacy class of $M_\Psi$ can be used to distinguish distinct coadjoint orbits. However, beware that this does not provide a complete classification: we will see that many orbits will correspond to any given $SL(2,\RR)$ conjugacy class. But for our purposes (in particular for determining the value of $\Delta$ labelling $T_{n,\Delta}$ classes), this will be sufficient

The conjugacy classes of $M_\Psi\in SL(2,\RR)$ fall into four distinct categories: identity, elliptic, hyperbolic, and parabolic. By choice of basis, we can put $M_\Psi$ into one of the following standard forms (real Jordan normal form):
\begin{equation*}
    M_I = \pm \begin{pmatrix}
        1 & 0\\
        0 & 1
    \end{pmatrix}, \quad M_E(\theta) = \pm\begin{pmatrix}
        \cos\tfrac{\alpha}{2} & -\sin\tfrac{\alpha}{2}\\
        \sin\tfrac{\alpha}{2} & \cos\tfrac{\alpha}{2}
    \end{pmatrix},\quad M_H(\Delta) = \pm \begin{pmatrix}
        e^{\frac{1}{2}\Delta} & 0\\
        0 & e^{-\frac{1}{2}\Delta}
    \end{pmatrix}, \quad M_P = \pm \begin{pmatrix}
        1 & 1\\
        0 & 1
    \end{pmatrix}\,,
\end{equation*}
where in elliptic/hyperbolic cases we have a single real parameter $0<\alpha<\pi$ or $\Delta>0$.

We can immediately classify the orbits with constant $b$ representatives in this way by solving \eqref{eq:Hill}. The $SL^{(n)}(2,\RR)$ orbits (with representatives $b=-\frac{c}{24}n^2$) give the identity conjugacy class with $M = (-1)^n I$, so both solutions to \eqref{eq:Hill} are (anti-)periodic. The remaining $U(1)_b$ orbits with $b<0$, $b>0$ and $b=0$ give elliptic, hyperbolic and parabolic conjugacy classes respectively.

To relate this classification to the above discussion, we note that the solutions to the equation \eqref{eq:invariance} expressing invariance of $b$ under the vector field $u$ are given by  products $u(\theta) = \psi_1(\theta)\psi_2(\theta)$ of two solutions to the Hill equation \eqref{eq:Hill}. In other words, $u(\theta)$ is given by a quadratic form in the solution space (a symmetric matrix $U$ once we have chosen a basis of solutions), and imposing periodicity implies that $U$ must be invariant under the monodromy matrix $M_\Psi$:
\begin{equation}
    u(\theta) = \Psi(\theta)^t U \Psi(\theta), \qquad M_\Psi^t U M_\Psi = U.
\end{equation}
For the identity conjugacy class containing $M_I$, any symmetric $U$ is invariant so we have a three-dimensional space of possible $u(\theta)$ (as we found for the $SL^{(n)}(2,\RR)$ orbits). In all other cases there is a unique $U$ up to rescaling, giving a one-dimensional space of possible $u$ (as for all other orbits).

We can also invert this, giving a basis of local solutions $\psi_\pm(\theta)$ in terms of $u(\theta)$ and the parameter $\lambda$ in \eqref{eq:bfromu}:
\begin{equation}\label{eq:Uinv}
    \psi_\pm(\theta) = \sqrt{|u(\theta)|} \;\exp\left(\pm \frac{\lambda}{2}\int^\theta \frac{1}{u}\right)\,.
\end{equation}
This applies either for $\lambda>0$ or for pure imaginary $\lambda$ (in which case a real basis of solutions is given instead by sines and cosines). For $\lambda=0$, a basis of solutions is instead given by $\sqrt{|u|}$ and $\sqrt{|u|}\int\frac{1}{u}$.

If $u$ is never zero, then any value of $\lambda$ (real, imaginary or zero) is allowed and we recover the $U(1)_b$ orbits (or $SL^{(n)}(2,\RR)$ when $\lambda \int\frac{1}{u} = 2\pi i n$).

If $u$ has any simple zeroes, then it is only possible to continue the solutions \eqref{eq:Uinv} smoothly if $|u'|=\lambda>0$ at all zeroes. In that case we can use a principal value prescription for the integral, and also perhaps pick up a  sign as we pass the zero: $\psi_+$ should flip sign when passing a zero where $u'>0$, while $\psi_-$ should flip sign at a zero where $u'<0$. If $\int\frac{1}{u}\neq 0$, this gives monodromy $M_H(\Delta)$ with \eqref{eq:Delta} (and the sign $(-1)^n$ with $2n$ zeroes). If $\int\frac{1}{u}= 0$ then get identity monodromy $M_I$ (corresponding to an $SL^{(n)}(2,\RR)$ orbit). Finally, if $u$ has higher-order zeroes then they must be double zeroes, and the special case $\lambda=0$ basis $\sqrt{|u|}$ and $\sqrt{|u|}\int\frac{1}{u}$ can be smoothly continued (with appropriate sign flips at the zeroes): this gives identity or parabolic monodromy $M_I$ or $M_P$ depending on the invariant $\varepsilon\in\{-,0,+\}$ computed from $\int\frac{1}{u}$ as discussed in the above section.

The point of this for us will be that a non-smooth choice of diffeomorphism $\phi$ along with the transformation rule \eqref{eq:Hilltransform} manifestly retains the invariance of the monodromy. This will allow us to classify orbits in the presence of certain singularities.

\section{Schwarzian theories}

\subsection{Classification of Schwarzian theories}

We are now in a position to describe the action of the most general Schwarzian theory. A theory is specified by a choice of coadjoint orbit, which we can specify by a representative $b(\theta)$ (along with a constant $c\neq 0$), and an adjoint vector (a function $u(\theta)$). The action is given as in \eqref{eq:Iabstract} by pairing $u$ with a point on the coadjoint orbit, which we can write using \eqref{eq:coAdjaction}:
\begin{equation}\label{eq:SchI2}
    I_{u,b}[\phi] = \int \frac{d\theta}{2\pi} \, u(\theta) \left[\phi'(\theta)^2 b(\phi(\theta))-\frac{c}{12}\Sch(\phi,\theta)\right], \qquad \phi \in \Diff/G_b \;.
\end{equation}
In this description, the dynamical variable of the theory can be taken as the diffeomorphism $\phi$, modulo the stabiliser group $G_b$ which leaves the action invariant. The dependence on $c$ is trivial, since it can be absorbed by rescaling $u$ and $b$.

For a fixed orbit we  think of $u$ as a `coupling function', and we would like to evaluate the integral over the orbit as a function of $u$:
\begin{equation}\label{eq:Psibu}
    \Psi_b[u] = \int \frac{\mathcal{D}\phi}{G_b} e^{i I_{u,b}[\phi]}.
\end{equation}
By construction, this integral depends only on the orbit $\orbit_b$, and not on which representative $b\in \orbit_b$ we choose.  So, the possible theories are classified by coadjoint orbits and hence fall into one of the families above. That is, we have the usual family of constant $b$ `elliptic' theories (with special  $SL^{(n)}(2,\RR)$ theories at $b=-\frac{c}{24}n^2$), the `hyperbolic' families $T_{n,\Delta}$ with $\Delta>0$ (meeting the $SL^{(n)}(2,\RR)$ theories at $\Delta=0$), or the special `parabolic' $\tilde{T}_{n,\pm}$. 

Note that $I_{u,b}$ is only bounded from below in the case of orbits with constant $b\geq -\frac{c}{24}$ representatives, and furthermore with couplings $u(\theta)$ which are positive everywhere: this exhausts all previously considered cases. So an integral of $e^{-I}$ makes sense only for those theories.  These theories arise in JT gravity with negative curvature, in Euclidean signature at an asymptotic spatial boundary. In this case,  the $S^1$ is interpreted as a thermal circle, so $\theta=\frac{2\pi}{\beta}\tau$ parametrises imaginary time $\tau$. The $b= -\frac{c}{24}$ orbit appears at the boundary of the hyperbolic disc (Euclidean AdS$_2$); cases with $-\frac{c}{24}<b<0$ correspond to putting a conical defect in AdS$_2$ \cite{Mertens:2019tcm} (while $b<-\frac{c}{24}$ would represent a conical excess); the $b>0$ theory appears at the boundary of a hyperbolic trumpet \cite{Saad:2019lba}.

Nonetheless,  integral weighted by $e^{i I}$ should make sense (as a distribution) in all cases. This is supported by considering Lorentzian de Sitter JT gravity in the next section, which will give us a physical interpretation of all of these theories.

\subsection{Anomalous  symmetry under conjugation of $u$}

The wavefunction $\Psi_b$ does not depend non-trivially on all the data about the $u(\theta)$, but only on its conjugacy class in $\Diff$. This is clear from the abstract definition of the action \eqref{eq:Iabstract} as the pairing of $u$ with the coadjoint $b$, which is left invariant by a simultaneous adjoint action on $u$ and coadjoint action on $b$. But here we have to be a little careful because of the central extension (the Schwarzian term in \eqref{eq:Adv}), giving us an anomalous transformation
\begin{equation}\label{eq:anom}
    \Psi_b\left[\frac{u\circ\phi}{\phi'}\right] = \exp\left(-i\frac{c}{12} \int \frac{d\theta}{2\pi}\frac{u(\phi(\theta))}{\phi'(\theta)} \Sch(\phi,\theta)\right)\Psi_b\left[u\right].
\end{equation}
This result has previously appeared for the standard Schwarzian theories (in the case when $u$ is positive everywhere) \cite{Cotler:2023eza,Stanford:2017thb}. Using this result, for a fixed theory we can learn everything by evaluating $\Psi_b$ on a single representative $u$ of each $\Diff$ conjugacy class. Alternatively, we can use this result as a consistency check.

\section{Schwarzian theories from dS$_2$ geometries}\label{sec:JT}

In this section we explain how the complete classification of Schwarzian theories appears in de Sitter JT gravity. For additional background on the interpretation of the wavefunction $\Psi^\text{WDW}$ and other aspects of the quantisation, see \cite{Held:2024rmg}.\footnote{Also see the group-theoretic perspective in \cite{Alonso-Monsalve:2025lvt}, which meshes very nicely with the classification of positive curvature geometries and Virasoro coadjoint orbits. However, this quantisation does not give direct access to Wheeler-DeWitt wavefunctions, the Schwarzian theories, and related observables at $\scri_+$.}

\subsection{Phase space and Wheeler-DeWitt wavefunctions of de Sitter JT gravity}

JT gravity is a theory of 2D gravity with a dilaton field $\Phi$. One way that this theory arises is from the dimensional reduction of 4D Einstein gravity on $S^2$, describing near-Nariai geometries \cite{Maldacena:2019cbz}. The Nariai spacetime is simply $dS_2\times S^2$ where $S^2$ has a fixed radius; JT gravity describes small fluctuations away from the (locally) $dS_2\times S^2$ geometry, with the dilaton field $\Phi$ giving the variation in the size of $S^2$.

The dilaton appears linearly in the action so it acts as a Lagrange multiplier, imposing the constant curvature condition $\mathcal{R}=2$. There is an interesting moduli space $\mathcal{M}$ of such constant curvature Lorentzian 2D geometries, which (almost)\footnote{The only mismatch is the branch of `big crunch' geometries indicated by the dashed line in the figure, which does not contain any asymptotic future infinity $\scri_+$.} matches the space of Virasoro coadjoint orbits as shown in figure~\ref{fig:moduli}.  This moduli space is the configuration space of pure JT gravity (i.e., the classical phase space of the theory is the cotangent bundle $T^*\mathcal{M}$, up to subtleties at the vertices) \cite{Held:2024rmg,Alonso-Monsalve:2024oii}. This means that a general quantum state is a superposition over such geometries: a wavefunction defined on $\mathcal{M}$.

To make the connection to the Schwarzian theory,  we consider the computation of a Wheeler-De-Witt wavefunction close to future infinity $\scri_+$. A wavefunction $\Psi^\mathrm{WDW}[a,\Phi]$ which solves the Wheeler-DeWitt equation can be constructed from a path integral over spacetimes ending on a Cauchy surface $\Sigma$, where we fix intrinsic data as a boundary condition. For dilaton gravity, this intrinsic data on $\Sigma$ (after fixing spatial diffeomorphisms so that the coordinate $\theta\sim\theta+2\pi$ is proportional to proper length) is the radius $a$ of the circle $S^1$ (so the proper length of $\Sigma$ is $2\pi a$), and the value of the dilaton $\Phi(\theta)$ as a function of the angle. In a state where the spacetime geometry $g$ is fixed (i.e., a wavefunction which is a delta-function on $\mathcal{M}$), this integral reduces to an integral over all possible Cauchy surfaces $\Sigma$ embedded in $g$ of length $\ell(\Sigma) = 2\pi a$, and in JT gravity the action reduces to a boundary term $\frac{1}{2\pi}\int ds \mathcal{K}\Phi$ where $\mathcal{K}$ is the extrinsic curvature of $\Sigma$ (so in particular, $\Psi$ depends only on local data near the embedded surface $\Sigma$). Schematically, the resulting path integral takes the form
\begin{equation}
    \Psi_g[a,\Phi] = \int_{\Sigma\subset g} \mathcal{D}\Sigma \,\delta(\ell(\Sigma)-2\pi a) e^{i \int \frac{d\theta}{2\pi} a \mathcal{K}(\theta)\Phi(\theta)},
\end{equation}
where $\mathcal{K}$ is the extrinsic curvature of $\Sigma$.

 \subsection{The Schwarzian limit for $T_{n,\Delta}$}

For general finite $a$, we expect this path integral to be difficult to calculate (similarly to AdS JT gravity at finite cutoff; see comments in \cite{Held:2024rmg}). But it simplifies in a limit $a\to\infty$, where $\Sigma$ approaches future infinity $\scri_+$ (or past infinity, which contributes a complex conjugate wavefunction). We will see that the interesting limit is when we simultaneously take $a\to\infty$ and  $\Phi(\theta)\to\infty$, holding fixed the ratio $u(\theta) = a^{-1}\Phi(\theta)$.

We will focus on the particular case corresponding to the interesting new $T_{n,\Delta}$ families of Schwarzian theories. For these, the relevant geometry takes the form
\begin{equation}\label{eq:DeltaMetric}
    ds^2 = -dt^2 + n^2 \cosh^2 t \,d\vartheta^2, \qquad (t,\vartheta=0)\sim (t-\Delta,\vartheta=2\pi),
\end{equation}
parametrised by the single modulus $\Delta>0$. The metric \eqref{eq:DeltaMetric} has a one-parameter non-compact isometry group generated by 
\begin{equation}\label{eq:isometry}
    \xi = \cos(n\vartheta)\partial_t -\frac{1}{n}\sin(n\vartheta) \partial_\vartheta.
\end{equation}
This gives us another useful way to characterise the spacetime: we start with the metric \eqref{eq:DeltaMetric} with $t,\vartheta\in\RR$ and quotient by the isometry which combines the flow of \eqref{eq:isometry} by parameter $\Delta$ with a shift $\vartheta\to\vartheta+2\pi$.

Given this spacetime, there is a one-parameter space of classical solutions for the dilaton (also invariant under the isometry \eqref{eq:isometry}), given by
\begin{equation}\label{eq:dilaton}
    \Phi = \varphi \cosh t \sin(n\vartheta),
\end{equation}
and $\varphi\in\RR$ is the coordinate on the fibre of the cotangent space $T_g^*\mathcal{M}$. Already at the level of this classical dilaton solution, we can see a qualitative difference with the previous appearance of the Schwarzian in JT gravity (AdS or dS).   As we approach $\scri_+$ (taking $t\to\infty$), the dilaton $\Phi$ blows up to $+\infty$ in some regions and $-\infty$ in others. In the reduction from 4D this is interpreted as a Schwarzschild-dS$_4$ black hole geometry, with the regions where $\Phi\to +\infty$ becoming four-dimensional parts of $\scri_+$ and the regions where $\Phi\to-\infty$ becoming black hole singularities. On any Cauchy surface, the renormalised dilaton $u$ will inevitably have $n$ fluctuations between positive and negative values. This is the JT perspective explaining why it is essential to allow the coupling function $u(\theta)$ to change sign.

We want to integrate over Cauchy surfaces near $\scri_+$ in this spacetime (i.e., at large $t$) with proper length $2\pi a \gg 1$, parametrised by a $2\pi$ periodic angle $\theta$ which we take to be proportional to the proper length (so the intrinsic metric is $a^2 d\theta^2$). The embedding of this slice is (locally) specified by a function $\theta \mapsto \vartheta(\theta)$ with $\vartheta'>0$. The $t$-coordinate $t(\theta)$ is then determined by the condition $ds^2 = a^2 d\theta^2$ on the intrinsic metric. Once we impose the requirement that $\Sigma$ stays near  $\scri_+$ (which is ultimately necessary to form a closed surface), this determines $t(\theta)$ uniquely in a large $a$ expansion:
\begin{equation}
    t(\theta) = \log\left(\frac{2a}{n \vartheta'(\theta)}\right) + O(a^{-2}).
\end{equation}
Globally, for $\Sigma$ to form a closed surface we need $\vartheta$ to satisfy a quasiperiodic boundary condition:
\begin{equation}\label{eq:quasibc}
    \tan\left(\tfrac{n}{2}\vartheta(\theta+2\pi)\right) = e^{\Delta} \tan\left(\tfrac{n}{2}\vartheta(\theta)\right).
\end{equation}
Additionally, to ensure that our surface $\Sigma$ winds once around the spacetime, the function $f(\theta) = \tan\left(\tfrac{n}{2}\vartheta(\theta)\right)$  must have $n$ simple poles in each $2\pi$ period, and must be increasing ($f'$ strictly positive) between the poles (so it also has $n$ simple zeroes per period). This follows from the description of the spacetime following \eqref{eq:isometry}: the isometry $\exp(\Delta\xi)$ acts near $\scri_+$ as a simple multiplication by $e^\Delta$ in the coordinate $\tan\left(\tfrac{n}{2}\vartheta\right)$, leading to \eqref{eq:quasibc} (see \cite{Held:2024rmg} for more details). This means that our integral over Cauchy surfaces can be expressed as an integral over increasing functions $\vartheta(\theta)$ obeying the quasiperiodic boundary conditions \eqref{eq:quasibc} and the discrete `single winding' condition of $n$ poles and zeroes per period  (modulo a gauge group which we comment on later).

The Schwarzian action appears when we expand the extrinsic curvature at large $a$:
\begin{gather}\label{eq:JTKb}
    \mathcal{K} = 1 + a^{-2} b_\vartheta(\theta) + \cdots, \\
    b_\vartheta(\theta) = -\frac{n^2}{2}\vartheta'(\theta)^2 - \Sch(\vartheta,\theta).
\end{gather}
The $1$ in the extrinsic curvature gives us a fixed large phase $a \int \frac{d\theta}{2\pi} \Phi(\theta)$, independent of the embedding of $\Sigma$ (in the AdS context, this would normally be subtracted by a counterterm). But the subleading term in the expansion gives a finite action $\int \frac{d\theta}{2\pi} u(\theta) b_\vartheta $ in the limit. This looks a lot like the Schwarzian action \eqref{eq:SchI}, with $c=12$ and $b = -\frac{1}{2}n^2$, corresponding to the $SL^{(n)}(2,\RR)$ orbits. And this is indeed precisely what we get for $\Delta=0$ when \eqref{eq:quasibc} becomes ordinary periodicity. But for $\Delta>0$ we must not forget to take into account the quasiperiodic boundary condition!

To connect with our classification of Schwarzian theories, we would like to rephrase this in terms of an ordinary diffeomorphism $\phi$ of $S^1$ with standard periodic boundary conditions. To do this, we simply choose any representative $\vartheta=\vartheta_0$ satisfying the desired boundary condition \eqref{eq:quasibc} and then write a general $\vartheta = \theta_0\circ\phi$ by composing with a standard periodic diffeomorphism $\phi$. If we write  $f(\phi) = \tan\left(\tfrac{n}{2}\vartheta(\phi)\right)$ as above, the action becomes the usual Schwarzian action \eqref{eq:SchI} for $\phi$, with $b(\phi) = -\Sch(f,\phi)$ and $c=12$. This means that the wavefunction is computed by one of our Schwarzian theories: the only remaining task is to identify which orbit $b$ belongs to.

For this, following the classification of section \ref{ssec:classify} we look for vector fields $u$ which leave $b$ invariant. If we make the ansatz $u(\theta)=\frac{g(f(\theta))}{f'(\theta)}$, we find that $b$ is invariant under the action generated by $u$ if $g'''(f)=0$. But we also require $u$ to be periodic, and since  $f(\theta+2\pi) = e^{\Delta} f(\theta)$ this only allows $g(f)\propto f$ for nonzero $\Delta$. So (up to normalisation) there is a unque solution for $u$, given by
\begin{equation}
    u(\theta) = \frac{f(\theta)}{f'(\theta)} = \frac{\sin(n\vartheta_0(\theta))}{n \vartheta_0'(\theta)}.
\end{equation}
At poles of $f$ (where $\vartheta$ is $\frac{\pi}{n}$ times an odd integer), we find $u=0$ and $u'=-1$, while at zeroes of $f$ (where $\vartheta$ is $\frac{\pi}{n}$ times an even integer)  we have $u=0$ and $u'=1$.  Since $f'$ (or $\vartheta_0'$) is positive everywhere else, $u$ is smooth and has no other zeroes. So, $u$ has $2n$ zeroes with $|u'|=1$ at each, as expected for the $T_{n,\Delta}$ orbit.  It remains only to calculate the invariant integral \eqref{eq:Delta} (with $\lambda=1$) to identify the orbit: 
\begin{equation}
    \int_{\mathrm{p.v.}} \frac{d\theta}{u(\theta)}  = \log\left(\frac{f(\theta+2\pi)}{f(\theta)}\right)=  \Delta.
\end{equation}
So the $\Delta$ we used to label the identification in \eqref{eq:DeltaMetric} indeed  corresponds precisely to the label $\Delta$ of the Virasoro coadjoint orbit $T_{n,\Delta}$ given in \eqref{eq:Delta}. This gives our final result\footnote{We have neglected a complex conjugate contribution from $\scri_-$ which perhaps should also be included, but whether this term is present depends on giving a more precise definition and interpretation to our wavefunction (see \cite{Held:2024rmg}).}
\begin{equation}
    \Psi^\text{WDW}_g[a,\Phi] \sim e^{i\int\frac{d\theta}{2\pi}a\Phi} \Psi^\text{Schw}_{n,\Delta}[u = \tfrac{\Phi}{a}],\qquad a,\Phi\to \infty.
\end{equation}

\subsection{The full classification}

The same ideas generalise very readily to other geometries and theories. Most of these (with $U(1)$ isometries) are extremely direct and straightforward. The only other tricky case is the parabolic theories. These can be obtained from a similar metric to \eqref{eq:DeltaMetric}, except where we identify using a different (parabolic) isometry generated by
\begin{equation}\label{eq:isopara}
    \xi = n\sin(n\vartheta)\partial_t +(1+\cos(n\vartheta)\tanh t) \partial_\vartheta.
\end{equation}
The corresponding dilaton solutions are $\Phi\propto \cosh (t) \cos (n \vartheta )+\sinh (t)$. The isometry $\xi$ in \eqref{eq:isopara} has a simple action on the coordinate $\tan\left(\tfrac{n}{2}\vartheta\right)$ and corresponding quasiperiodicity to \eqref{eq:quasibc}, but now by adding a constant rather than multiplying by a constant. This leads to the $\tilde{T}_{n,\pm}$ theories, with the label $\pm$ depending on the sign of the shift.

The last point we must address is the gauge group, corresponding to the stabiliser of $b$. This is the one-dimensional non-compact subgroup group of diffeomorphisms generated by $u$. It corresponds precisely to shifting the surface by the isometry  $\xi$ given in \eqref{eq:isometry}. Of course, this shift will not change the action and will give a zero mode of the path integral. To get a finite wavefunction, we must exclude this mode by hand (or formally, divide by its volume). This is equivalent to the procedure required to define a sensible group-averaging inner product when configurations preserve a non-compact subgroup of the gauge symmetry. The unbroken symmetry group can depend on the state, in which case states with different symmetry groups belong to different superselection sectors of the theory with incomparable relative normalisations of the inner product \cite{Marolf:1995cn,Marolf:2008hg} (and the discussion of \cite{Held:2025mai} mostly in the context of higher-dimensional de Sitter space). The importance of this  for JT gravity was emphasised in \cite{Alonso-Monsalve:2025lvt}.

\section{Solving the new Schwarzian theories}\label{sec:solution}

\subsection{Smooth classical solutions}\label{ssec:classSols}

The first step towards solving the Schwarzian theories is to understand the classical solutions.

As commented in section \ref{ssec:coadjtheories} , classical solutions on the coadjoint orbit $W$ correspond precisely to points $b\in W$ which are invariant under the coadjoint action of the coupling $u$, $\ad_u^* b=0$. We already characterised these in our classification of coadjoint orbits in section \ref{ssec:classify}, where we showed that for a given $u(\theta)$ we can immediately write the solution $b(\theta)$ up to an undetermined constant $\lambda^2$ (which can be positive or negative). It then remains to find whether this $b$ lies on the orbit $W$ corresponding to the theory of interest: 
\begin{equation}\label{eq:ClassSol}
    \text{$b$ is classical solution for $\Psi_\orbit[u]$} \iff b = \frac{c}{24}\frac{2u u''-(u')^2+\lambda^2}{u^2} \in \orbit.
\end{equation}

For $u>0$ (or $u<0$), it is straightforward to identify the orbit $W$ corresponding to each such $b$ and hence to identify the classical solutions. There is a unique constant representative on the same orbit as $b$:
\begin{equation}
    b = \frac{c}{24}\frac{2u u''-(u')^2+\lambda^2}{u^2} \quad\xrightarrow{\quad\Ad^*\quad}\quad b = \frac{c}{24}\lambda^2\left( \int \frac{1}{u}\right)^2.
\end{equation}
So, for any of the $U(1)_b$ or $SL^{(n)}(2,\RR)$ orbits there is a unique choice of $\lambda^2$, and hence a unique classical solution for every such $u$. But for any of our novel orbits (in the hyperbolic $T_{n,\Delta}$ or parabolic $\tilde{T}_{n,\pm}$ classes), there is no solution when $u$ has no zeroes (and in particular, no solution when $u$ is a constant).

Since $u>0$ generates a $U(1)$ Hamiltonian flow on the orbit, the Duistermaat-Heckman theorem immediately applies and the result of our coadjoint orbit integral will be one-loop exact. For the $T_{n,\Delta}$ and $\tilde{T}_{n,\pm}$ where there are no classical solutions, this implication is that the Schwarzian path integral vanishes identically, $\Psi_b[u]=0$ for $u>0$.

When $u$ has zeroes, if we are searching for smooth solutions $b(\theta)$ then we are more constrained. Smoothness at zeroes of $u$ requires $\lambda^2 = (u')^2$, so $|u'| = \lambda$ must be equal at all zeroes. If this is satisfied with $\lambda>0$ and $u$ has $2n$ zeroes, then $b\in T_{n,\Delta}$ with $\Delta = \lambda \int\frac{1}{u}$, so we have a smooth classical solution in that orbit (but no others). In the special case $\int \frac{1}{u}=0$, this is a classical solution for the $SL^{(n)}(2,\RR)$ theory. If $\lambda=0$ then we have the additional constraints that $u''\neq 0$ and $u'''= 0$ at all the zeroes of $u$, and $b$ is in either a $\tilde{T}_{n,\pm}$ or $SL^{(n)}(2,\RR)$ orbit, where $n$ is the number of (double) zeroes (which one is determined from the sign of the regularised integral $\int\frac{1}{u}$ as described in section \ref{ssec:classify}). For the generic $U(1)_b$ orbits there is never a smooth solution when $u$ has zeroes.

 The upshot is that a smooth classical solution will only exist if $u$ is in a specific conjugacy class of vector fields under $\Diff$ (up to a rescaling), depending on the orbit in question. For the $T_{n,\Delta}$ and $\tilde{T}_{n,\pm}$ orbits this is quite constraining, with smooth solutions only existing on a submanifold of codimension 2 or higher in the space of coupling fields $u$. However, there is a reason that we have been careful to specify "smooth solution" throughout this section. We will soon be led to allow solutions with certain singularities, which will relax the conditions on $u$ for a solution to exist.

If we do have a solution, then we can immediately evaluate the on-shell action $I=\int u b$:
\begin{equation}\label{eq:Ionshell}
    I_\mathrm{on-shell}[u] = \frac{c}{24}\int \frac{d\theta}{2\pi} \frac{\lambda^2-(u')^2}{u},
\end{equation}
where we have dropped a total derivative term. It is straightforward to check that under diffeomorphisms this on-shell action shifts precisely by the anomaly \eqref{eq:anom}.

\subsection{Setting up the one-loop calculation}\label{ssec:1loopsetup}

Now we would like to calculate the one-loop integral around a smooth solution $b(\theta)$ for a theory with coupling $u(\theta)$. This takes the form of a Gaussian functional integral over functions $v(\theta)$ which generate diffeomorphisms:
\begin{equation}\label{eq:Psi1loop}
    \Psi_\text{one-loop}[u] = e^{i I_\mathrm{on-shell}[u]} \int \frac{\mathcal D v(\theta)}{\mathrm{Stab}(b)} \Pf' \Omega \exp\left[\frac{i}{2}\int \frac{d\theta}{2\pi} v(\theta) Q v(\theta)\right] = e^{i I_\mathrm{on-shell}[u]}\sqrt{\frac{\det' \Omega}{\det' Q}}.
\end{equation}
The integral excludes the zero mode $v\propto u$ (and two additional modes for the $SL^{(n)}(2,\RR)$ theories)  generating the stabilizer of our classical solution $b$.

We have a measure factor given by the Pfaffian $\Pf' \Omega = \sqrt{\det' \Omega}$ of the symplectic form $\Omega_b$ evaluated at the classical solution, where the prime reminds us to exclude the zero mode $u$. We can write this form in terms of the third order differential operator
\begin{equation}
    \Omega v = \frac{c}{12}v''' -2 b v' - b'v\,,
\end{equation}
so that $\Omega_b(v_1,v_2) =  \int \frac{d\theta}{2\pi} v_1\Omega v_2$. The Pfaffian can also be written as a path integral over a fermionic function $\psi(\theta)$ with action $\frac{1}{2}\Omega_b(\psi,\psi)$.

Finally, $Q$ is the operator giving the quadratic action. For a general coadjoint orbit, we can expand the action at the group element $g=\exp(v)$ as
\begin{equation}\label{eq:Iexpansion}
\begin{aligned}
    I[g] &= \langle b, \Ad_{g^{-1}} u\rangle \\
    &= \langle b,u\rangle - \langle b,[v,u]\rangle + \tfrac{1}{2}\langle b,[v,[v,u]]\rangle + \cdots \\
    & = I_\mathrm{on-shell} + \tfrac{1}{2} \Omega_b(v,[v,u]) + \cdots \,,
\end{aligned}
\end{equation}
where in the last line we have used the fact that we're expanding around a classical solution to drop the linear term. From this, we can write the quadratic fluctuation operator as $Q = \Omega A = -A^\dag \Omega$, where $Av = -[u,v]$ is the adjoint action of $u$ on $v$. For our case, we can drop the central term in $[u,v]$ (since it does not contribute) to write
\begin{equation}\label{eq:QOmegaA}
    Q = \Omega A, \qquad A v = uv'-v u'= u^2 \left(\frac{v}{u}\right)'\,.
\end{equation}
The resulting $Q$ is a fourth order differential operator. 

To write \eqref{eq:Psi1loop} in the form of a ratio of determinants, we have implicitly chosen a measure for the integral $\mathcal{D}v$, namely the flat measure with respect to the coordinate $\theta$. Explicitly, this means that we are defining the determinants by regarding $\Omega$ and $Q$ as (anti-)self adjoint operators on the Hilbert space $L^2(S^1,d\theta)$, diagonalising these operators, and taking a (regularised) product over their eigenvalues. This choice of flat measure is arbitrary since it depends on the choice of coordinate, but the ratio of determinants in the final result should not depend on this choice (we could instead use $\mu(\theta)d\theta$ for any positive measure $\mu$ on $S^1$).\footnote{A natural choice is the diff-invariant measure $u(\theta)^{-3}d\theta$. This simplifies the calculation for the case $u>0$, but is not applicable in other cases since it is not positive. The positive alternative $|u(\theta)|^{-3}d\theta$ is also a poor choice since the singularities at zeroes of $u$ make physically sensible smooth modes non-normalisable, including the gauge mode $u$.} This will be verified by the result being diff-invariant (as expressed in \eqref{eq:anom}). It is important to be explicit about this definition of the measure, because we will encounter some subtleties which require careful treatment of technical aspects.

From the relation $Q = \Omega A $, it is tempting to write a relation like $\det Q = \det \Omega \det A$ and reduce the one-loop calculation to a determinant of the first-order operator $A$, without ever introducing the auxiliary Hilbert space structure. For the case $u>0$, attacking the problem directly in this way works straightforwardly (as we show in appendix \ref{app:alt1loop}). But for the general case, it is not so easy to address the technical subtleties using this approach: an ad hoc prescription reproduces our final answer, but we do not know how to justify this prescription in a more principled way. Nonetheless, our calculation will be guided by a more precise and careful version of this idea.

\subsection{One loop calculation for $U(1)$ orbits}

We first implement this one-loop calculation of $\Psi_b[u]$ for the $U(1)$ orbits (excluding the $SL^{(n)}(2,\RR)$ special cases), for which existence of a classical solution requires that $u(\theta)$ is never zero; we will take $u>0$. The full calculation we present here is not really necessary, since in this case we can change to coordinates where the classical solution $b$ is constant (using \eqref{eq:anom}), making direct calculation very straightforward. But we will not have this luxury for other orbits, so it will be a helpful warm-up to show a method that generalises to such cases, without the distraction of the additional subtleties which we will then encounter.

In these cases, it is straightforward to define $\Omega$ and $Q$ as operators on the Hilbert space $L^2(S^1)$: they are essentially (anti-)self-adjoint, with no need to specify boundary conditions. We will describe this in more detail when we turn to other cases, since it will then fail for $Q$! Additionally,  $\Omega$ and $Q$ each have the single zero mode $v\propto u$. To show this for $Q$, note that $Q v=\Omega A v=0$ requires $A v\propto u$, which has solution $v = c_1 u + c_2 u \int \frac{1}{u} $; this is periodic only if $c_2=0$.

So, it remains only to calculate the ratio of determinants of these operators as in \eqref{eq:Psi1loop}. We will not calculate the determinant directly, but instead the variation of the ratio of determinants as we change $u$ and/or the orbit in question. We can write the variation as
\begin{equation}\label{eq:deltalogdet}
    \delta \log \left(\frac{\det'\Omega}{\det'Q}\right) = \Tr(\delta \Omega \, G_\Omega) - \Tr(\delta Q \, G_Q),
\end{equation}
where the Green's function $G_\Omega$ (resp.~$G_Q$) is the inverse of $\Omega$ (resp.~$Q$) on the space orthogonal to the zero mode $u$. So we can write
\begin{equation}
    \Omega G_\Omega = G_\Omega \Omega = Q G_Q = G_Q Q = 1-P,
\end{equation}
where $P$ is the orthogonal projection onto $u$. Making use of $Q = \Omega A$, we will find an expression for $G_Q$ using $G_\Omega$, in such a way that the $\Omega$-dependent terms cancel.

To construct $G_Q$, we want to find an operator which maps $f$ to a solution $v$ of $Q v = \Omega A v = f$ whenever $f$ is orthogonal to the zero mode $u$. This means that $A v = G_Q f +k u$ for some constant $k$. So to find $v$ from $g = G_Qf$ we need a Green's function $G_A$ which inverts $A$ modulo a multiple of $u$,
\begin{equation}
    v = G_A g \implies Av -g  = k\, u \,.
\end{equation}
The proportionality constant $k$ is fixed uniquely by requiring $g+ku$ to be in the image of $u$, which is the orthogonal complement to $u(\theta)^{-2}$ (the kernel of the adjoint $A^\dag$). Explicitly, this gives $k = \int u^{-2}g / \int u^{-1}$. Restating this, we can write the defining property of $G_A$ as the identity
\begin{equation}\label{eq:AGA}
    A G_A = 1 - \hat{P}, \qquad \hat{P} =  \frac{|u\rangle\langle u^{-2}|}{\langle u^{-2}|u\rangle}.
\end{equation}
This definition does not determine $v = G_A f$ uniquely, but only up to a multiple of $u$ (so $G_A$ is determined up to addition of $|u\rangle\langle \psi|$ for any function $\psi(\theta)$). We remove this final ambiguity by orthogonal projection $1-P$ (to ensure that our final result is orthogonal to $u$). This projection could be absorbed into $G_A$, but it will be convenient to  leave it explicit and correspondingly allow some freedom in the choice of $G_A$. Putting this together, we have
\begin{equation}\label{eq:GQ}
    G_Q = (1-P)G_A G_\Omega.
\end{equation}

An explicit Green's function which satisfies these requirements (found by solving $Av = g+c u$ with $g(\theta) = \delta(\theta-\theta')$) is
\begin{equation}\label{eq:GA}
    G_A(\theta,\theta') = -\left(\textstyle\int \frac{1}{u}\right)^{-1} \frac{u(\theta)}{u(\theta')^2} \int_{\theta'}^\theta \frac{1}{u} ,\qquad \theta'<\theta<\theta+2\pi\,,
\end{equation}
and any other $G_A$ will differ by $u(\theta)$ times an arbitrary function of  $\theta'$. This satisfies
\begin{equation}
    \begin{gathered}
    A G_A (\theta,\theta') = \delta(\theta-\theta') -  \left(\textstyle\int \frac{1}{u}\right)^{-1} \frac{u(\theta)}{u(\theta')^2},
    \end{gathered}
\end{equation}
which is simply a more explicit form of \eqref{eq:AGA}. This expression for $G_A$ will turn out to be convenient because it commutes with $A$, also satisfying $G_A A = 1-\hat{P}$ (which we can check by acting with $A^\dag$ on $G_A(\theta,\theta')$ as a function of $\theta'$). Additionally, it satisfies $G_A|u\rangle = 0$ (so $G_A P = 0$).

Now we have everything we need to calculate the variation in \eqref{eq:deltalogdet}:
\begin{align*}
    \Tr(\delta \Omega  G_\Omega) - \Tr(\delta Q (1-P)G_A G_\Omega) &= \Tr(\delta \Omega  G_\Omega) - \Tr(\Omega \delta A G_A G_\Omega) - \Tr(\delta \Omega A G_A G_\Omega) - \Tr(Q \delta P G_A G_\Omega)
    \\
    &= - \Tr(\delta A G_A )  + \Tr (G_\Omega\delta \Omega \hat{P}) - \Tr(\delta P(1-\hat{P})) \\
    &= - \Tr(\delta A G_A ) 
\end{align*}
In the first line we have used the variation of $QP=0$ to replace $\delta Q P$ with $-Q \delta P$. The second line simply uses cyclicity of $\Tr$ and identities for $G_\Omega\Omega$, $G_A A$, $A G_A$, and $G_A P=0$. In the final line we find a cancellation making use of the variations of the identities $\Omega \hat{P}=0$ and $(1-P)\hat{P} = 0$, as well as $\Tr \delta P = 0$ (since $\Tr P=1$).

So, we end up finding that we need only compute $\Tr(\delta A G_A)$, the na\"ively expected result from the cancellation of $\det' \Omega$ using $Q = \Omega A$. We have
\begin{equation}
    \delta A \, G_A(\theta,\theta') = \frac{\delta u(\theta)}{u(\theta)}  \delta(\theta-\theta') -\left(\textstyle\int \frac{1}{u}\right)^{-1} \frac{\delta u(\theta)u'(\theta) - \delta u'(\theta)u(\theta)}{u(\theta')^2} \int_{\theta'}^\theta \frac{1}{u} -\left(\textstyle\int \frac{1}{u}\right)^{-1} \frac{\delta u(\theta)}{u(\theta')^2} .
\end{equation}
To take the trace, we set $\theta'\to \theta $ and integrate. The first term gives a divergence; we regulate simply by subtracting this. For the second term, the limit is ambiguous depending on whether $\theta'$ approaches $\theta$ from left or right (which means setting $\theta' = \theta-2\pi$ in the above formula). But in any case, the resulting integral takes a $\theta$-independent value, so this term will vanish in the trace since the prefactor becomes a total derivative (of $\delta u/u$). So the only interesting contribution is the final term, which gives us the result
\begin{equation}
    \Tr(\delta A G_A) = \delta \log \int \frac{1}{u}.
\end{equation}

So, our final result (up to an overall regulator-dependent constant) for the integral over the $U(1)$ orbit with constant representative $b$ and arbitrary $u(\theta)>0$ is
\begin{equation}
    \Psi_{U(1),b}[u] = \left(\int \frac{d\theta}{u(\theta)}\right)^{-1/2} \exp\left[2\pi i b \left(\int \frac{d\theta}{u(\theta)}\right)^{-1} - i \frac{c}{24}\int \frac{d\theta}{2\pi} \frac{(u')^2}{u}\right].
\end{equation}
This correctly reproduces the already-known result.

\subsection{$T_{n,\Delta}$ at one loop}\label{ssec:hyponeloop}

Now, let's see what happens when we try to generalise the above one-loop calculation to the case where $u(\theta)$ has zeroes. Specifically, consider expanding around a smooth classical solution $b$ of the $T_{n,\Delta}$ orbit. This means that $u$ has $2n$ distinct zeroes $\theta_k$, with $u'(\theta_k) = (-1)^k \lambda>0$, and $\lambda\int\frac{1}{u} = \Delta$, which we take to be positive (so in particular, the principal value integral $\int\frac{1}{u}$ round the circle is non-zero). The solution $b$ is given by \eqref{eq:ClassSol}.

A straightforward generalisation of the previous subsection fails for a technical reason: the quadratic fluctuation operator $Q$ is not essentially self-adjoint! This happens because the eigenvalue equation $Q v = \mu v$ has a singular points at $\theta_k$, where the coefficient $u$ of the highest derivative term  in $Q$ vanishes. We find that this equation is underspecified: we need to choose boundary conditions at each $\theta_k$ in order to have a good eigenvalue problem with real spectrum of eigenvalues $\mu$ as befits a self-adjoint operator. If we do not impose any boundary conditions, then the spectrum contains all $\mu \in \CC$, and we have no hope of making sense of the one-loop determinant.

We defer a complete discussion and analysis (including a review of the theory of self-adjoint extensions) to appendix \ref{app:ESA}. Here we will summarise the main points and state the results we need. The pertinent properties of $Q$ follow from the leading-order behaviour of the highest derivative terms near a zero of $u$ (which we place at $\theta=0$):
\begin{equation}
    Q v \sim  \theta v'''' + 2v''' +\cdots.
\end{equation}
Near $\theta=0$, the local  solutions to the eigenvalue equation $Q v=\mu v$ (or to the inhomogeneous equation $Qv = f$ with smooth source $f$) have an expansion
\begin{equation}\label{eq:domQ}
    v(\theta) \sim a_0 + a_1\theta +a_1'\theta\log|\theta| +a_2\theta^2 + \cdots,
\end{equation}
where the coefficients $a_i$ can be chosen independently. Requiring $v$ to be a weak (distributional) solution to the equation imposes only three boundary conditions (for a fourth-order ODE) relating the values of $a_i$ in the regions $\theta<0$ and $\theta>0$, namely that $a_0$, $a_1'$ and $a_2$ must take the same value on each side. There is no such condition on $a_1$: a function $\theta \Theta(\theta)+\cdots$ (where $\theta$ is the step function) with a discontinuous first derivative is a valid distributional solution at $\theta=0$. To make $Q$ into a uniquely defined self-adjoint operator, we need one additional boundary condition at $\theta=0$.

For now we will simply state the boundary condition which we will take as our definition of the theory, deferring some justification of this choice (using input from de Sitter JT gravity) until the next section. We make the natural choice that the coefficient $a_1$ \eqref{eq:domQ} must also take the same value as $\theta\to0$ from positive or negative values. This allows for non-smooth functions of the $\theta\log|\theta|$ singularities, but forbids jumps in the derivative (in the sense that $v'(\epsilon)-v'(-\epsilon)$ must go to zero as $\epsilon\to 0$). This defines a unique self-adjoint extension of $Q$, and hence a well-defined functional determinant.

Given this choice of self-adjoint extension, $Q$ does not have any physical zero modes (i.e., with the exception of the `gauge mode' $u$). Locally, a basis of solutions to $Qv=0$ is given by
\begin{equation}\label{eq:QZM}
    Qv = 0 \implies v(\theta) = u(\theta)\left(c_0 + c_1 {\textstyle\int^\theta}\tfrac{1}{u} + c_+ \exp\left(\lambda{\textstyle\int^\theta}\tfrac{1}{u}\right)+ c_- \exp\left(-\lambda{\textstyle\int^\theta}\tfrac{1}{u}\right)\right).
\end{equation}
To impose our boundary conditions at a zero of $u$ (where $u'=\pm\lambda$), we can simply define $\int^\theta\frac{1}{u}$ with a principle value prescription. Since $\int_{S_1}\frac{1}{u}\neq 0$, only the  solutions with $c_1=c_+=c_-=0$ are periodic.

There is one other natural choice of boundary conditions (the only other choice which is local and respects $\theta\to-\theta$ reflection symmetry): this disallows $\theta\log|\theta|$ terms (setting $a_1'=0$ in \eqref{eq:domQ}) but allows discontinuous first derivatives. This choice also appears to give a sensible theory, at least at the level of classical solutions and one loop fluctuations. This alternative definition has qualitatively different behaviour  due to the appearance of additional zero modes. We comment on this in section \ref{ssec:altbcs}.

Now we have given a definition for $Q$ (and established that it has no physical zero modes), we can go ahead and compute its determinant to evaluate the $T_{n,\Delta}$ path integral $\Psi_\Delta[u]$ at one loop. Fortunately, we can immediately reuse the calculation of the preceding section for the $U(1)$ orbits! Though to do this, we have to take care to correctly deal with the zeroes of $u$. In particular, we need to check that $G_Q$ as defined in \eqref{eq:GQ} is indeed an inverse for the operator $Q$ with the correct boundary conditions, which requires that the image of $G_Q$ should be in the domain of $Q$ (with the expansion as in \eqref{eq:domQ}). This holds if we take the integral in \eqref{eq:GA} to be defined with a principal value prescription. We should similarly take care with singularities such as those arising in the $u(\theta')^{-2}$ denominator in \eqref{eq:GA}; once again, a principle value definition takes care of these (where it's important that $G_A$ only needs to act on sufficiently smooth functions). The upshot is that we can import the same result for the determinant, only adding the proviso of the principle value:
\begin{equation}
    \Psi_{n,\Delta}[u] = \left(\int_\pv \frac{d\theta}{u(\theta)}\right)^{-1/2} \exp\left[i\frac{c}{24}\int \frac{d\theta}{2\pi} \frac{\lambda^2-(u')^2}{u}\right],
\end{equation}
where (so far) this result applies for the case where $u$ has $2n$ zeroes, with $|u'|=\lambda$ at all of these. Once again, the overall constant is regulator dependent (though a reasonable argument could be made for a fixed phase by counting the number of negative modes of $Q$).

\subsection{Singular solutions}\label{ssec:singsols}

In the previous section, we found that a good definition of the one-loop determinant required us to consider fluctuations $v(\theta)$ which are not smooth, allowing singularities of the form $\theta \log|\theta|$. But this has other consequences once we allow finite diffeomorphisms generated by such vector fields, and consider the orbit of coadjoints $b(\theta)$ under the resulting diffs.

We can see this from the classical and linearised one-loop analysis, by noticing that two facts are in apparent conflict. First, smooth classical solutions only exist on a submanifold of the coupling functions $u(\theta)$ of codimension $2n$ (we have the conditions that $|u'|=\lambda$ at each zero, where $\lambda\int\frac{1}{u} = \Delta$). Second, the quadratic fluctuation operator $Q$ has no physical zero modes, so is invertible on the tangent space of the coadjoint orbit. These are in tension, because the linearised equations of motion under a variation $\delta u(\theta)$ take the form $Q v = \Omega \delta u$; if $Q$ is invertible, then this has solutions for arbitrary (smooth) $\delta u$. The resolution is that $Q$ is invertible only once we allow $v$ to have singularities of the form $\theta \log|\theta|$ at zeroes of $u$, and this results in correspondingly singular changes to the classical solution which go like $\delta b  = \Omega v \sim \frac{1}{\theta^2}$.

To go beyond this linearised analysis, we can look at the finite diffeomorphisms generated by exponentiation a singular vector field. We can consider a case where $v = \theta\log|\theta|$ in some finite interval containing $\theta=0$:
\begin{equation}\label{eq:singdiff}
    v(\theta) = \theta\log|\theta| \xrightarrow{\quad\phi= \exp(s v)\quad} \phi(\theta) = \sgn(\theta)|\theta|^p \qquad (p = e^s>0)\,.
\end{equation}
So, by exponentiating vector fields $v(\theta)\sim\theta\log|\theta|$ we allow diffeomorphisms which go as an arbitrary positive power. If we apply a coadjoint action under such a diff, we get a singularity from the Schwarzian derivative:
\begin{equation}
    \Sch(\phi,\theta) \sim \frac{1-p^2}{2\theta^2}.
\end{equation}
The upshot is that we allow classical solutions of the form \eqref{eq:ClassSol} which we have already given, but without the constraint on derivatives at zeroes from smoothness. We only have the inequality constraint that $\lambda^2>0$. The action \eqref{eq:Ionshell} is also unchanged, except that the singularities mean that the integral must be defined using a principle value prescription (which can be motivated as the only prescription which respects $\theta \to -\theta$ symmetry, or alternatively from JT gravity as in section \ref{sec:JTnonsmooth}).

Note that the vector field $u$ which stabilises $b$ remains smooth under such a diff: if $u(\theta) = \theta$ in a neighbourhood of $\theta=0$, then the adjoint action (under $\phi^{-1}$) maps this to $\frac{u(\phi(\theta))}{\phi'(\theta)} = \frac{\theta}{p}$. This changes $u'(0)$ as expected, but does not give rise to any singularities in $u$. It also leaves the principal value integral $\int_{S_1} \frac{1}{u}$ unchanged. So, relaxing our definition of diffs to allow non-smooth $\phi$ of this form collapses the conjugacy classes of vector fields $u(\theta)$ with $2n$ simple zeroes; the only remaining invariant is $\int \frac{1}{u}$.

A consequence is that we can no longer classify coadjoint orbits using the conjugacy class of the stabilizer vector field $u$ as in \ref{ssec:classify}. We might even worry that the classification itself changes (as it did for vector fields) once we enlarge the space of allowed diffeomorphisms! Fortunately, the alternative method using the Hill equation goes through essentially unchanged. This does require us to specify the required behaviour of solutions to the Hill equation at singularities, but this can be deduced from applying the weight $-1/2$ transformation law in \eqref{eq:Hilltransform} to our singular diffs. In the simplest case where we start with $b=0$ in an interval (and corresponding smooth solutions $\psi(\theta)= A + B\theta$), applying  $\phi(\theta) = \sgn(\theta)|\theta|^p$ with $p>0$ results in
\begin{equation}
    \psi''(\theta) -\frac{p^2-1}{4\theta^2}\psi(\theta)=0, \qquad \psi(\theta)= A|\theta|^\frac{1-p}{2}   + B\sgn(\theta) |\theta|^\frac{1+p}{2}\, .
\end{equation}
The general case is similar (matching this result in an  expansion as $\theta\to 0$).

With this prescription (derived from the transformation law for $\psi$), it is manifest that the monodromy of the Hill equation remains invariant under our more general class of diffeomorphisms. In particular, the solutions given in \eqref{eq:Uinv} (along with similar comments following that equation about changes in sign) continue to be a valid basis of solutions to the Hill equation, which diagonalise the monodromy. So, the formula $\Delta = \lambda \int \frac{1}{u}$ remains true.

The upshot is that for every $u(\theta)$ with $2n$ simple zeroes and $\int_\pv \frac{1}{u} \neq 0$, there is exactly one representative of the  $ T_{n,\Delta}$ coadjoint orbit extended by our non-smooth diffs which is invariant under the action of $u$:
\begin{equation}\label{eq:bSing}
    b = \frac{c}{24}\frac{2u u''-(u')^2+\lambda^2}{u^2}\in T_{n,\Delta} \iff \text{$u$ has $2n$ simple zeroes},\quad \Delta = \lambda\int\frac{1}{u}>0.
\end{equation}
In other words, after allowing for this class of singular solutions, there is a unique classical solution for the $ T_{n,\Delta}$ theory for every such $u$ (and still no solution otherwise).

We could re-analyse the one-loop integrals in the case that $b$ has singularities. But this is unnecessary, since we can (by a diff with singularities) use the transformation rule \eqref{eq:anom} to reduce to the previous case. This gives our final result for the $T_{n,\Delta}$ path integral:
\begin{equation}\label{eq:psinDelta}
    \Psi_{n,\Delta}[u] = \left(\int_\pv \frac{d\theta}{u(\theta)}\right)^{-1/2} \exp\left[i\frac{c}{24}\left(\frac{\Delta^2}{2\pi} \left(\int_\pv \frac{d\theta}{u(\theta)}\right)^{-1} -\int_\pv \frac{d\theta}{2\pi} \frac{(u')^2}{u}\right)\right],
\end{equation}
where $u$ has exactly $2n$ simple zeroes.

We can verify this by checking the one-loop fluctuations around a singular solution $b$. This requires us to redo the analysis of self-adjoint extensions, now for $\Omega$ as well as $Q$, which we do in appendix \ref{app:ESA}. We find that this works straightforwardly for $0<p<\frac{3}{2}$.

However, we encounter a problem in the range of parameters $p\geq \frac{3}{2}$, since one of the local solutions to the eigenvalue equation ($v\sim |\theta-\theta_0|^{1-p}$) becomes non-normalisable in the $L^2$ inner product. In this case, our strategy of defining the one-loop calculation via determinants of self-adjoint operators $\Omega$ and $Q$ on $L^2(S^1)$ ceases to be applicable. We nonetheless find it more compelling to continue to define the theory through the invariance \eqref{eq:anom} (including under our non-smooth diffeomorphisms) and hence the result \eqref{eq:psinDelta} (for example, we will see no barrier for $p>3/2$ in JT gravity in section \ref{sec:JTnonsmooth}). This motivates finding an alternative way to define and compute the one-loop path integral without the auxiliary Hilbert space $L^2(S^1)$ (using only the symplectic form $\Omega_b(v_1,v_2)$ and the quadratic form $Q(v,v)=\int vQv$): we would be interested to know of a principled method that achieves this aim.

\subsection{The parabolic $\tilde{T}_{n,\pm}$ theories}

Once we allow for our class of non-smooth diffeomorphisms and singular solutions $b$, we can almost immediately deduce a similar result for parabolic orbits  in the case of $u$ with only simple zeroes.

If we take one of our solutions for a $T_{n,\Delta}$ theory given in \eqref{eq:bSing} (constructed from $u$ with $2n$ simple zeroes and $\int\frac{1}{u}>0$) and simply set $\lambda=0$, then we get a solution $b(\theta)$ with parabolic monodromy for the Hill equation \eqref{eq:Hill}: the two independent solutions are $\psi(\theta) = \sqrt{u(\theta)}$ and $\psi(\theta) =  \sqrt{u(\theta)} \int^\theta \frac{1}{u}$. The sign classifying the monodromy (which was previously used to distinguish between $\tilde{T}_{n,+}$ and  $\tilde{T}_{n,-}$ classes) is simply the sign of $\int\frac{1}{u}$. So, the path integral for the parabolic theories $\tilde{T}_{n,\pm}$ is obtained simply as a $\Delta\to 0$ limit of the $T_{n,\Delta}$ theories, with a restriction on the sign of the coupling:
\begin{equation}\label{eq:psiparabolic}
    \Psi_{n,\pm}[u] = \left(\int_\pv \frac{d\theta}{u(\theta)}\right)^{-1/2} \exp\left[-i\frac{c}{24}\int_\pv \frac{d\theta}{2\pi} \frac{(u')^2}{u}\right] \; \Theta\left(\pm \int\frac{1}{u}\right).
\end{equation}
This seems to be the reasonable definition of the parabolic $\tilde{T}_{n,\Delta}$ theories which is compatible with our definition of the hyperbolic $T_{n,\Delta}$ theories.

It should be pointed out that the singular classical solution considered above is not strictly in the orbit of the smooth classical solutions that we described earlier! Even allowing for isolated singularities in our diffeomorphisms $\phi$, the number of intervals of a vector field $u$ where $u>0$ and $u<0$ must remain invariant (because the Jacobian $\phi'(\theta)$ can never be negative). Nonetheless, the smooth classical solutions $b$ where $u$ has double zeroes can be obtained as a limit of our singular solutions where simple zeroes coalesce, and the principle value integral $\int\frac{1}{u}$ is finite in this limit.\footnote{To see this, it is sufficient to work locally and consider the family $u(\theta) = \theta^2-\epsilon^2$; the principle value integral $\int \frac{1}{u}$ over the interval $\theta\in (-a,a)$ is $-2a^{-1} + O(\epsilon^2)$ in the limit $\epsilon\to 0^+$.}

However, we do not reach the same conclusion if we try to directly compute the one-loop integral expanded around the smooth classical solution, using the philosophy employed above (via the auxiliary Hilbert space $L^2(S^1)$). The problem is essentially the same as that encountered in the final paragraph of the above section for $p\geq \frac{3}{2}$. Here, the quadratic fluctuation operator $Q$ is essentially self-adjoint but requires boundary conditions which are not obtained as a limit of those described in section \ref{ssec:hyponeloop}. The reason is that the  singular $v\sim \theta\log|\theta|$ solutions which we have allowed morph into $v\sim \frac{1}{\theta}$ when zeroes coalesce, which is non-normalisable in the $L^2$ norm and hence must be excluded (see appendix \ref{app:ESA} for details). So instead, we are forced to choose boundary conditions which are analogous to the alternative boundary conditions which we comment on in section \ref{ssec:altbcs}. We can recover the result \eqref{eq:psiparabolic} by working over a different function space which allows $\theta^{-1}$ singularities, but it would be nice to understand this in a more principled and less ad hoc manner.

\subsection{The $SL^{(n)}(2,\RR)$ theories}

The final class of theories we must return to is the $SL^{(n)}(2,\RR)$ Schwarzian, including the original Schwarzian theory for $n=1$ (which appears at $\scri^+$ in de Sitter space dS$_2$). While the result for $u>0$ (or $u<0$) is already known and straightforward to obtain, we would like to evaluate the path integral $\Psi[u]$ in the case that $u$ has zeroes. So, for the remainder of this section we will always consider $u$ with $2n$ simple zeroes.

The result of this theory are qualitatively different from anything we have already encountered. We can see this already by looking at classical solutions, including the case that $b$ has double pole singularities. We saw already that if the principle value integral $\int\frac{1}{u}$ is non-zero, then the corresponding classical solutions given in \eqref{eq:bSing} will always lie in one of the hyperbolic $T_{n,\Delta}$ orbits (for $\lambda=0$) or a parabolic $\tilde{T}_{n,\pm}$ orbit (for $\lambda=0$). This means that there can only be classical solutions for $SL^{(n)}(2,\RR)$ theories --- even singular solutions --- if $\int\frac{1}{u}=0$.

Let's see how this restriction arises from the computation of the one-loop determinant expanded around a smooth classical solution.\footnote{Note that in this case, there is a very simple and obvious representative of the orbit which can be used to make the analysis more concrete: namely, $u \propto \sin(n\theta)$ (which indeed satisfies $\int\frac{1}{u}=0$)and constant $b = -\frac{c}{24}n^2$.} Once again, the quadratic fluctuation operator is not essentially self-adjoint, and we impose the same boundary condition at the zeroes of $u$ to define a self-adjoint extension. But a qualitative difference from previous cases appears when we consider zero modes of $Q$. We wrote the explicit solutions for $Qv=0$ in \eqref{eq:QZM}: all four solutions take the form of $u(\theta)$ times a function of $\int^\theta\frac{1}{u}$. The boundary conditions at zeroes of $u$ are satisfied if we define this integral with a principle value prescription. But now, since $\int_{S_1}\frac{1}{u}=0$, all four modes are zero modes! Three of these are now gauge modes (also zero modes of $\Omega$) which we do not integrate over, but the fourth zero mode, $v_0(\theta) =u(\theta) \int^\theta\frac{1}{u}$, is a genuine physical zero mode.

To understand the implications of this zero mode for the path integral, consider what happens when we slightly change the coupling to $u+\delta u$. Now the linear term in the expansion of the action \eqref{eq:Iexpansion} combines with the shift in the coupling to give a term in the action which is linear in $v$:
\begin{equation}\label{eq:sl21loop}
    \Psi[u+\delta u] = e^{i\langle b,\delta u\rangle}\int \frac{\mathcal{D}v}{\mathfrak{sl}(2)} \Pf'(\Omega) \exp\left[i \int\frac{d\theta}{2\pi}\left(\tfrac{1}{2} vQv + \delta u \Omega v \right)\right].
\end{equation}
For invertible  $Q$ (with no physical zero modes) this this could simply be absorbed by a shift of $v$ (moving to a slightly shifted classical solution), at the cost of a change in the on-shell action which is quadratic in $\delta u$ (combing with the linear change in the prefactor). But if we have a zero mode $v_0$, integration over this mode (at least at one loop) imposes a delta function constraint, $\Psi \propto \delta(\int \delta u \Omega v_0)$. For our zero mode $v_0$, we have $\Omega v_0 \propto \frac{1}{u^2}$, so the argument of the delta function is $\int \frac{\delta u}{u^2}$. But this is precisely the variation $\delta \int\frac{1}{u}$, so the delta function enforces that we lie on the submanifold of couplings which permit a classical solution.

It is now unnecessary to compute the remaining one-loop determinant, since there are simply no remaining parameters that the result can depend on (the result is just a regulator-dependent number). This allows us to write down our final result for the path integral:\footnote{A first sight the scaling of the one-loop factor with $u$ looks strange, since the na\"ive expectation is that multiplying $u$ by a factor of $g{-2}$ should rescale the prefactor in $\Psi[u]$ by a factor of $g^{-3}$ (as in the usual Schwarzian path integral result, line 2 of table \ref{tab:results}). I'd like to thank Douglas Stanford for pointing this out. The argument for this is that the quadratic fluctuation operator $Q$ scales by $g^{-2}$, which can be absorbed by scaling the integration variable $v$ by a factor of $g$ at a cost of $g^{-3}$ in the integration measure (where $3$ is the number of gauge modes). The resolution of this puzzle is that the argument of the delta function should be thought of as its linearised version $\int \frac{\delta u}{u^2}$, and for the one-loop scaling argument to work we must rescale $\delta u$ by a factor of $g^{-1}$ (rather than $g^{-2}$) to leave the integrand in \eqref{eq:sl21loop} invariant.}
\begin{equation}\label{eq:SL2result}
    \Psi_{SL^{(n)}(2,\RR)}[u] = \delta\left(\int_\pv \frac{d\theta}{u(\theta)}\right) \exp\left(-i\frac{c}{24}\int \frac{d\theta}{2\pi} \frac{(u')^2}{u}\right),\qquad \text{$u$ has $2n$ simple zeroes}.
\end{equation}
This completes the solution of the full menagerie of Schwarzian theories for all possible coupling functions $u$, summarised in table~\ref{tab:results}.

\subsection{An alternative boundary condition}\label{ssec:altbcs}

We now comment on the solution to an alternative definition of the theory, defined by choosing the other natural boundary condition giving a self-adjoint extension of the quadratic fluctuation operator $Q$. This condition excludes $\log$ singularities of the form $v(\theta)\sim\theta\log|\theta|$, but allows $v'(\theta)$ to be discontinuous at zeroes of $u$. This choice gives qualitatively different results for the path integral: we will indicate the main features, though we do not solve these alternative theories in full detail.

We will discuss mostly the $T_{n,\Delta}$ theories.  We begin by considering the one-loop calculation expanded around a smooth classical solution, meaning that we have a coupling $u(\theta)$ with $2n$ zeroes $\theta_k$ where $u'(\theta_k) = \pm \lambda$.  The alternative boundary conditions admit $2n$ zero modes, $v_k(\theta) = \id_{[\theta_k,\theta_{k+1}]}(\theta) u(\theta)$, where $\id_I$ denotes the characteristic function taking the value $1$ on the interval $I$ and zero elsewhere. In other words, these zero modes are $u$ on an interval between two zeroes of $u$, and zero elsewhere. One linear combination of these modes is the usual gauge mode $u$, leaving $2n-1$ physical zero modes, which impose $2n-1$ constraints. Following the logic of the previous subsection, the path integral contains a product of delta functions with the following arguments:
\begin{equation}\label{eq:altZM}
    \int\delta u\Omega v_k\propto \left(u'(\theta_{k+1})\delta u'(\theta_{k+1}) - u''(\theta_{k+1})\delta u (\theta_{k+1})\right) - \left(u'(\theta_k)\delta u'(\theta_k) - u''(\theta_k)\delta u (\theta_k)\right).
\end{equation}
Now, the two terms can also be written as $\tfrac{1}{2}\delta(u'(\theta_k)^2)$, where the second term accounts for the shift $\delta \theta_k = -\frac{\delta u(\theta_k)}{u'(\theta_k)}$ of the zero. So, the $2n-1$ zero modes impose the constraint that $|u'(\theta_k)|$ is equal at all zeroes $\theta_k$ of the coupling function. 

This constraint is precisely what it required to get a solution of the form given in \eqref{eq:ClassSol} without singularities $b\sim \frac{1}{(\theta-\theta_k)^2}$ (by choosing $\lambda = |u'(\theta_k)|$). However, there is no constraint fixing $\lambda \int\frac{1}{u}=\Delta$ as would be required for a smooth $b(\theta)$ to remain a member of the correct orbit.\footnote{There is a would-be zero mode $u\int\frac{1}{u}$ (with any choice of integration constants in each interval) which would impose precisely such a constraint. But this is excluded by the alternative boundary conditions due to the $\theta\log|\theta|$ singularities. Hypothetically, one might waste a great deal of time trying to make sense of a theory which also allows this zero mode. This is quite attractive since in such a theory, the path integral would impose $2n$ delta-function constraints, which are precisely the constraints that $u$ is in the conjugacy class of a smooth solution for the given coadjoint orbit. However, such an attempt seems doomed to failure, since we are inevitably left with a non-self-adjoint quadratic fluctuation operator $Q$, and so it seems to be impossible to make sense of the spectrum and one-loop determinant. But perhaps an alternative perspective makes sense of such a theory.} So how can we get a solution for any value of $\Delta$ without constraining $\lambda \int\frac{1}{u}$? The answer is that $b$ has a different kind of singularity. We can deduce what this looks like (at least linear order near a classical solution) from the action of $\Omega$ on the zero modes $v_k$ (essentially the same calculation as \eqref{eq:altZM}): the new singularities take the form $b(\theta)\propto \delta'(\theta-\theta_k) +\frac{u''(\theta_k)}{u'(\theta_k)} \delta(\theta-\theta_k) $. This is a weak solution of the differential equation \eqref{eq:invariance} (regarded as an equation for $b$) giving the condition for invariance under $u$. Allowing for such singularities in $b$ at each of the $2n$ points $\theta_k$, we have a $2n$-dimensional space of solutions for given $u$. One of these dimensions corresponds to $\Delta$ (characterising the orbit to which $b$ belongs), and the remainder give $(2n-1)$-dimensional space of solutions for each $T_{n,\Delta}$ theory, which is the manifold of flat directions obtained by following the zero modes.

Beyond linear order, these singular coadjoints $b$ arise from the action of non-smooth diffeomorphisms that we get from exponentiating the vector fields $v_k$. These are continuous functions $\phi(\theta)$ from $S^1$ to $S^1$ with discontinuous derivatives at $\theta_k$: locally at $\theta_k=0$, perhaps $\phi(\theta) = A_-\theta$ for $\theta<0$ and $\phi(\theta) = A_+\theta$ for $\theta>0$ with $A\pm>0$.  The Schwarzian derivative $\Sch(\phi,\theta)$ of such diffs is not well-defined (though perhaps there is some regulated version that makes sense), so it does not appear that $b(\theta)$ has any sensible definition as a distribution. However, it does have a clear and unambiguous definition via a generalisation of the Hill equation as described in section \ref{ssec:classifyHill}. By starting with a smooth solution and using the transformation \eqref{eq:Hilltransform} under a singular diffeomorphism, we find a jump condition for the solutions $\psi(\theta)$ at the singular point $\theta=0$: $\psi(0_+) = \alpha \psi(\theta_-)$ and $\psi'(0_+) = \alpha^{-1} \psi'(\theta_-)$, where $\alpha^2 = \frac{A_+}{A_-}$. The Hill equation augmented with this jump condition serves as an indirect definition of singular $b$, and the monodromy of this equation (noting that the jump preserves $\det(M)=1$) can be used to classify the coadjoint orbit to which it belongs.

In summary, the result for the path integral for this version of the theory should take the form
\begin{equation}
    \Psi^{\text{alt}}_{n,\Delta} = \left(\prod_{k=1}^{2n-1} \delta(u'(\theta_{k+1})^2-u'(\theta_k)^2) \right)f\left(\int\frac{1}{u},\lambda \right)\,\exp\left[ i\frac{c}{24}\int \frac{d\theta}{2\pi} \frac{\lambda^2-(u')^2}{u}\right],
\end{equation}
where $u(\theta_k)=0$ and $u'(\theta_k)=\pm \lambda$. The function $f$, which must depend only on the two remaining invariants $\lambda$ and $\int\frac{1}{u}$, is determined by a one-loop calculation which we have not attempted to do.\footnote{In particular, it is not totally clear how to treat the measure for the zero modes in this calculation, since the symplectic form $\int v_{k} \Omega v_{k+1}$ on zero modes is not unambiguously defined.}

With these boundary conditions, the story is extremely similar for the  $SL^{(n)}(2,\RR)$ theories when $u$ has zeroes. The zero mode $u \int\frac{1}{u}$ that we encountered above is no longer admissible, so we do not enforce the $\int\frac{1}{u}=0$ constraint, but we do have the $2n-1$ zero modes which impose equality of  $|u'(\theta_k)|$.

For the parabolic $\tilde{T}_{n,\pm}$ theories, we start by expanding around a smooth solution, so the coupling $u$ must have $n$ double zeroes $\theta_k$ (and also $u'''(\theta_k)\neq 0$). As mentioned above (and explained in appendix \ref{app:ESA}), the quadratic fluctuation operator is essentially self-adjoint, and the natural alternative boundary condition definition of the theory takes this at face value. In this case, there are $n$ physical zero modes. As before, we have $n$ zero modes of the form $u(\theta)\id_{[\theta_k,\theta_{k+1}]}(\theta)$, one linear combination of which is the gauge mode $u(\theta)$. The last mode is $u(\theta)\int^\theta\frac{1}{u}$, where we can choose independent integration constants in each interval $(\theta_k,\theta_{k+1})$: the choices differ by linear combinations of the other zero modes. A convenient basis for the $n$ zero modes is to write them all as $v_k(\theta) = u(\theta)\int^\theta\frac{1}{u}$, and define the integral with a principle value prescription at all zeroes except $\theta_k$ (and an arbitrary overall integration constant). This is smooth everywhere except $\theta_k$ (using $u'''=0$ at zeroes to avoid $\log$ singularities), where it has a discontinuous second derivative. We find that $\Omega v_k \propto \delta(\theta-\theta_k)$, so under perturbations of the coupling integration over the mode $v_k$ enforces the constraint that $\delta u(\theta_k)=0$. This is the linearisation of the constraint that $\theta_k$ remains a double zero. So, the path integral for the parabolic theories with this definition should be proportional to a $\delta$-function constraint enforcing that $u$ has $n$ double zeroes, so that we have a smooth classical solution $b$.\footnote{Nothing seems to enforce $\delta u '''(\theta_k)=0$, so perhaps the $u'''=0$ constraint should be relaxed and the more mild $\frac{1}{\theta-\theta_k}$ singularities should be permitted in the solutions $b$.}

\section{Fermionic localisation}\label{sec:localisation}

So far we have only evaluated the path integral at one loop order. For the previously considered Schwarzian theories, it was argued \cite{Stanford:2017thb} that this is sufficient: the theory is one-loop exact by the Duistermaat-Heckman (DH) theorem \cite{Duistermaat:1982vw}. However, for our new theories  (and for the more general coupling functions $u(\theta)$ for the old theories), the same conclusion does not immediately follow, since a key assumption for the DH formula --- namely, that the action generates a $U(1)$ --- does not hold. In this section we argue that the conclusion nonetheless remains true. But to show this, we must go back to the fermionic localisation argument for the DH theorem, and fill in the gap left by the absence of $U(1)$ invariance.

Let's briefly review the fermionic localisation argument (following \cite{Stanford:2017thb}, which the reader can refer to for more details, though we use slightly different notation).  Let $M$ be a symplectic manifold with coordinates $x^i$ and symplectic form $\omega$ and let $H(x)$ be a Hamiltonian which generates a $U(1)$ flow $\zeta^i$. This means that solutions to $\dot{x}^i = \zeta^i= \omega^{ij}\partial_j H$ are periodic with a fixed period. We would like to evaluate the integral of $e^{H}$ over $M$ with the symplectic measure, which we can write in terms of an integral over fermions $\psi^i$:
\begin{equation}
    Z = \int d x d\psi \; e^{i H + \frac{1}{2}\omega_{ij}\psi^i\psi^j}.
\end{equation}
Now, the action is invariant under a fermionic symmetry
\begin{equation}
    \delta x^i = \psi^i,\quad \delta \psi^i = \zeta^i.
\end{equation}
Concretely, the fermions $\psi^i$ are the exterior algebra of forms $dx^i$, and the symmetry acts on differential forms by the exterior derivative plus interior multiplication by $\zeta$, $\delta = d+ \iota_\zeta$. From this, it's easy to see (from Cartan's magic formula) that the square of this symmetry is the generator of the flow, i.e.~the Lie derivative with respect to $V$, $\delta^2 = \mathcal{L}_\zeta$. To make use of this symmetry, note that deforming the action by the variation $\delta V$ of any $\zeta$-invariant quantity $V$ does not change the integral. To prove one-loop exactness, we can choose $V = \Gamma_{ij}\psi^i \zeta^j$ where $\Gamma_{ij}$ is any $\zeta$-invariant metric on $M$, with invariance required so that $\delta^2 V = 0$. This gives $\delta V = \Gamma_{ij}\zeta^i \zeta^j + (\text{fermions})$, so adding $\delta V$ to the action with a large coefficient will localise the path integral to points where $\zeta^i=0$, which are the classical solutions.

An essential ingredient in the above proof is that we need to find a $\zeta$-invariant metric $\Gamma$. For the Duistermaat-Heckman theorem, this is guaranteed by the fact that the flow generates a $U(1)$: an invariant metric can be constructed by choosing any metric whatsoever, and integrating (or averaging) that metric over the flow. Nonetheless, the conclusion still holds if we can explicitly find a $\zeta$-invariant metric $\Gamma$. That is what we will do below. Note also that when the deformation is weighted by $i$ in the exponent (as is natural for our `Lorentzian' integrals), it is sufficient for $\Gamma$ to be a non-degenerate symmetric form: it need not be positive-definite. The  integral is controlled at large deformation parameter by the stationary phase approximation.

We start by looking at the quadratic action linearised around a classical solution, and the corresponding linearised fermionic symmetry:
\begin{equation}
    I_\mathrm{quad}[v,\psi] = \frac{1}{2}\int\frac{d\theta}{2\pi}\left(vQv +\psi\Omega\psi\right), \qquad \delta v=\psi,\quad \delta \psi = Av.
\end{equation}
Here $v$ is the bosonic infinitesimal diffeomorphism and $\psi$ its fermionic partner; both should be regarded as functions of $\theta$ modulo multiples of the zero mode $u$. The operator $A$ is the adjoint action of $u$, defined explicitly in \eqref{eq:QOmegaA}. The action is invariant under $\delta$, using $Q = \Omega A$, and $\delta^2=A$.

In this context, an invariant metric is equivalent to a hermitian operator $\Gamma$ which satisfies $\Gamma A + A^\dag \Gamma=0$. Given such an operator, we can deform the action by adding a multiple of
\begin{equation}\label{eq:deltaLoc}
    \delta \int \psi\, \Gamma A v = \int \left(vA^\dag \Gamma A v + \psi \Gamma A \psi\right).
\end{equation}
This satisfies $\delta^2 = 0 $ if $\Gamma A +A^\dag \Gamma = 0$, so in that case the path integral is invariant. The simplest possible such operator $\Gamma$ is just a multiplication operator: a function of $\theta$. The invariance condition then becomes a differential equation, which is solved by $\Gamma = u(\theta)^{-3}$. In the special case that $u$ is a constant, this reproduces the deformation considered in \cite{Stanford:2017thb}, but generalised to be a diff-invariant; when $u>0$, this metric gives the natural $u$-invariant (and positive-definite) pairing of vector fields $\Gamma(v_1,v_2) = \int \frac{v_1 v_2}{u^3}$.  When $u$ has zeroes this operator not positive-definite, but this does not present any trouble for the localisation argument in our context.

We should take some care to check that the singularities at zeroes of $u$ do not present any obstruction: in particular, we need to return to the original theory, including boundary conditions, when we take the deformation away. To see this, we can use the explicit form of the deformation \eqref{eq:deltaLoc}, noting that it is precisely equivalent to changing the background classical solution $b$ by adding a multiple of $u(x)^2$. In other words, the deformation is the same as a shift in the parameter $\lambda^2$ in \eqref{eq:ClassSol}. In the case that $u$ has simple zeroes, this gives us the one-loop determinant around one of the singular classical solutions discussed in section \ref{ssec:singsols} (and appendix \ref{app:ESA}). So the fact that our one-loop determinant \eqref{eq:psinDelta} for fixed $u$ is independent of $\Delta$ is explained as a consequence of the fermionic symmetry, and in fact must hold exactly. Given this, we can calculate the path integral in the $\Delta\to\infty$ limit where loops are suppressed to conclude that the result is one-loop exact.

\section{Singular solutions in JT gravity}\label{sec:JTnonsmooth}

We now return to de Sitter JT gravity, addressing the choice of boundary conditions in our definition of $Q$ and the associated singular configurations. To do this, we restore the cutoff $a$ giving the proper length of the Cauchy surface as a finite (but large) number. First, we look for classical solutions which reproduce the singular behaviour. We will find that this is not quite sufficient, but instead show that there are admissible off-shell configurations corresponding to the singular diffeomorphisms under consideration. We also explain why the alternative boundary conditions considered in \ref{ssec:altbcs} do not come from reasonable configurations in JT gravity.

\subsection{Classical solutions in JT with finite cutoff}

We are interested in the behaviour near a point on $\scri_+$ where the asymptotic dilaton changes sign. We can choose coordinates so that a classical spacetime solution locally takes the form
\begin{equation}\label{eq:flatdS}
    ds^2 = \frac{-d\eta^2+dx^2}{\eta^2},\qquad \Phi = -\lambda\frac{ x}{\eta} \, \qquad (\eta<0),
\end{equation}
with the conformal boundary $\scri_+$ at $\eta=0$. The constant $\lambda$ is related to a well-known invariant in JT gravity, $\Phi^2+(\nabla\Phi)^2 = \lambda^2$ (which is a constant in spacetime on classical solutions).

We first search for classical solutions for evaluation of the wavefunction $\Psi[a,\Phi]$ with $\Phi(\theta) = a u(\theta)$ (locally, without worrying here about extending to a global solution). A classical solution means an embedded Cauchy surface with specified intrinsic data,
\begin{equation}
    ds^2 = a^2 d\theta^2,\qquad \Phi(\theta) = a u(\theta).
\end{equation}
The extrinsic curvature of such a surface can be expressed simply in terms of the time coordinate $T(s)$ defined by $\eta = -e^{-T}$, as a function of proper length $s$:
\begin{equation}
    \mathcal{K}(s) = \frac{\ddot{T}(s)+\dot{T}(s)^2+1}{\sqrt{1+\dot{T}(s)^2}},  \qquad \eta = e^{-T}.
\end{equation}

In the Schwarzian limit we take $a\to\infty$ with $x(\theta)$ fixed and use the approximation $|\eta'|\ll |x'|$ to find
\begin{equation}
    \frac{x'(\theta)}{x(\theta)} \sim \frac{\lambda}{u(\theta)}, \qquad \mathcal{K} \sim 1-a^{-2}\Sch(x(\theta),\theta) + \cdots.
\end{equation}
Using this to express $\mathcal{K}$ in terms of $u(\theta)$ and relating to the coadjoint $b(\theta)$ as in \eqref{eq:JTKb}, we recover the classical solution \eqref{eq:ClassSol} (in particular, giving us a JT interpretation of the parameter $\lambda$).

A smooth Cauchy surface (parametrised so that $x(0)=0$) will always have a dilaton profile $\Phi\sim \lambda \theta$ for small $\theta$, so can be a classical solution only when $u(\theta)\sim \lambda\theta$. Unsurprisingly, this recovers the constraint that we found earlier for smooth solutions for the Schwarzian theory. But what if we relax the requirement of smoothness? Do we get solutions in JT (at finite $a$) with more mild singularities than those we found in the Schwarzian limit?

Using the dilaton value to eliminate $\eta$, we get a quadratic equation for $\frac{x'}{x}$:
\begin{equation}\label{eq:JTembedding}
    \frac{u(\theta)^2}{\lambda^2} \frac{x'(\theta)^2}{x(\theta)^2} - a^{-2}\left(\frac{x'(\theta)}{x(\theta)}-\frac{u'(\theta)}{u(\theta)}\right)^2 = 1.
\end{equation}
The Schwarzian approximation breaks down close to a zero of $u(\theta)$, and we have to retain higher-order terms in $a$ in that vicinity. The solution of \eqref{eq:JTembedding} which recovers the Schwarzian when $a\to\infty$ is
\begin{equation}
    \frac{x'(\theta)}{x(\theta)} = \frac{ a^2\lambda u(\theta)}{ a^2 u(\theta)^2-\lambda^2}\sqrt{1+\frac{u'(\theta)^2-\lambda^2}{a^2u(\theta)^2}} -  \frac{\lambda^2}{a^2 u(\theta)^2-\lambda^2}\frac{u'(\theta)}{u(\theta)}.
\end{equation}
Working near a zero, it suffices to treat $u$ in a linear approximation. To get rid of redundant parameters, we can change coordinate from $\theta$ to the proper distance $s$ along the curve (using dots to denote $s$ derivatives), and define the positive parameter $p = \frac{\lambda}{u'(0)}$. It is sufficient to study the region $s>0$ (with $s<0$ related by symmetry), and this gives
\begin{equation}\label{eq:xdx}
    \frac{\dot{x}(s)}{x(s)} = \frac{p \left(s+s^{-1}\right)}{p+s \sqrt{s^2+1-p^2}}.
\end{equation}
For large $s$ the solution is $x(s)\sim A\left(\frac{s}{a}\right)^p$ for a constant $A$, which should be of order unity to recover the usual Schwarzian limit away from the singularity. Comparing in this limit, the parameter $p$ indeed corresponds to $p$ in the singular diffeomorphisms in \eqref{eq:singdiff}.

Now we separate into two cases, starting with $p<1$. In this case, we have a solution by integrating \eqref{eq:xdx}, and this solution goes linearly as $x(s)\sim B a^{-p} s$ for small $s$ (with $B$ of order one). This gives us a perfectly good classical solution. It is not smooth at $s=0$: it will have a kink since $\eta$ is extended to $s<0$ as an even function of $s$, and $\dot{\eta}(s)$ is not continuous (meaning that the extrinsic curvature $\mathcal{K}$ has a $\delta$-function contribution there). But since $\Phi(s)$ vanishes at that point, such a kink is allowed for a classical (weak) solution, and gives an unambiguous finite action. The conclusion is that the singular classical solutions in the Schwarzian for $p<1$ get regulated to admissible classical solutions in JT gravity, justifying their inclusion in our analysis of the Schwarzian theory.

However, things are rather different for $p>1$: there is no real solution $x'/x$ once $\theta$ gets sufficiently small (when $a^2 u(\theta)^2< \lambda^2(p^{-2}-1)$)! For such a singularity, there is simply no classical solution in JT corresponding to the singular classical solution of the Schwarzian theory.

At this point one might be tempted to throw out this range of parameters and try to make sense of a theory which allows only singularities of the $0<p<1$ type. But this is too quick. The reason is that there is no regime where the classical approximation is controlled in the region where $\Phi$ is small: quantum effects are never negligible. So, it is quite possible to have configurations in the path integral with which are classically forbidden. So, we look for such off-shell configurations next.\footnote{There are complex classical solutions for $p>1$, so perhaps these control the evaluation of the path integral at large but finite $a$ in analogy to tunnelling instantons. But these are not straightforward to interpret, and would ideally be understood as saddle-points of a real path integral over configurations such as those discussed in the following section.}

\subsection{A Cauchy surface from singular diffeomorphisms}

For an alternative perspective, we will not simply look for classical solutions (which are not reliable in any case close to the singularity, where the effective value of $\hbar$ goes to infinity). Instead, we look for off-shell configurations which correspond to non-smooth diffeomorphisms $\phi(\theta) \sim \theta^p$ (for $\theta>0$). We will be happy to include these if they describe a sensible Cauchy surface, and  if the Lagrangian $ \mathcal{K}\Phi$ remains finite so that we have a finite, unambiguous action.\footnote{Integrable singularities would also be OK, but we will in fact find a finite Lagrangian.}

Working in our coordinates \eqref{eq:flatdS}, we are looking for an embedded Cauchy surface $(\eta(\theta),x(\theta))$ satisfying
\begin{equation}\label{eq:embeddiff}
    x(\theta) = \phi(\theta), \qquad \frac{-\eta'(\theta)^2+\phi'(\theta)^2}{\eta(\theta)^2} = a^2.
\end{equation}
This gives us a differential equation for $\eta$, which we must solve with the condition that we recover the expected Schwarzian solution $\eta \sim -a^{-1}\phi'$ away from the singularity (i.e., when $\theta - O(1)$ and $a\gg 1$). To study this, we make the ansatz
\begin{equation}\label{eq:JTdiff}
    \eta(\theta) = - a^{-1} \phi'(\theta)e^{-f(\theta)} \implies \left(f'(\theta) -  \frac{\phi''(\theta)}{\phi'(\theta)}\right)^2 = a^2\left(e^{2 f(\theta)}-1\right).
\end{equation}
To recover the Schwarzian, we require $f \ll 1$ and $f'\ll 1$, from which we can immediately deduce
\begin{equation}
    f(\theta)\sim \frac{1}{2a^2}\left(\frac{\phi''(\theta)}{\phi'(\theta)}\right)^2 +\cdots.
\end{equation}
 It is straightforward to systematically correct this order-by-order in powers of $a^{-2}$ (with $f'$  determined at each order by the terms at previous orders). If $f(\theta)\sim\theta^p$, we have $f(\theta)\sim \frac{(p-1)^2}{2a^2\theta^2}$, and this approximation will be valid when $a\theta\gg 1$.

It will be important to understand the stability of this solution to small perturbations. Linearising $f$, we find
\begin{equation}
    \frac{\delta f'(\theta)}{\delta f(\theta)} \sim -a^2 \frac{\phi'(\theta)}{\phi''(\theta)}.
\end{equation}
This grows rapidly in the direction of increasing $\theta$ when $\phi''<0$ (which is the case for $0<p<1$), meaning that we expect the requirement $f\to 0$ when $\theta\gg a^{-1}$ to pick out a unique allowed solution. On the other hand, when $\phi''>0$ (the case for $p>1$), the perturbation decays at increasing $\theta$, so any perturbation at very small $\theta$ will continue to match the Schwarzian solution. The choice of solution for $f$ will instead be determined by what happens at $\theta\to 0$ in this case.

We now zoom into the region of the singularity $\theta\ll 1$, choosing $\phi(\theta) = \theta^p$ there. It is convenient to change coordinate from $\theta$ to proper length $s=a\theta$ (abusing notation by using the same name for the function $f$), where we have
\begin{equation}
    \left(f'(s)-\frac{p-1}{s}\right)^2 = e^{2f(s)}-1,
\end{equation}
and we want to match onto a solution with $f(s)\sim \frac{(p-1)^2}{2s^2}$ as $s\to\infty$. From the above stability analysis, we expect this $s\to\infty$ boundary condition to select a unique solution for $0<p<1$, but not for $p>1$. Since these two cases are qualitatively different, we study them separately.

\subsubsection*{The case $0<p<1$}

In this case, to match the correct large $s$ behaviour we are solving $f'(s) =  \sqrt{e^{2f(s)}-1} -\frac{1-p}{s}$. Numerical studies support the prediction that there is a unique real solution which goes to zero at large $s$ (instead of blowing up $f\to\infty$ or hitting the singular point $f=0$). We would like to see what this solution looks like at small $s$.

There is a self-consistent approximation for which $f$ blows up logarithmically as $s\to 0$, of the form
\begin{equation}
    f(s)\sim (1-p)\log\frac{A}{s} + \frac{A}{p} \left(\frac{s}{A}\right)^p + \cdots
\end{equation}
for some constant $A$ (of order one). We verified numerically that this indeed describes the small $s$ behaviour of the exact solution. Translating back to the function $\eta(\theta)$, we find the expansion
\begin{equation}
    \eta\sim -p a^{-p}A^{1-p} + \theta^p + \cdots,
\end{equation}
which we can also verify this directly from \eqref{eq:embeddiff} by assuming that $a\eta\ll 1$ (which requires $(\eta')^2\sim(\phi')^2$ to make the denominator small). This configuration is not identical to the classical solution we found above, but the parametric scaling of $\eta(0)$ as $a^{-p}$ is the same.

If we now compute the extrinsic curvature at small $s$, we find $\mathcal{K}(s)\sim -\frac{1-p}{s}$.\footnote{A distributional definition the singularity at $s=0$ could be given using a Lorentzian Gauss-Bonnet theorem to define $\int\mathcal{K}$, but this is unnecessary for us except to demonstrate that $\int\Phi\mathcal{K}$ does not have a singular contribution from $s=0$.} This is milder than the $\theta^{-2}$ singularity of $b(\theta)$ in the Schwarzian limit, and in particular makes the Lagrangian $\mathcal{K}(s)\Phi(s)$ bounded.  In the $a\to\infty$ limit, the resulting action reproduces the principle value prescription used in our solution to the Schwarzian theory (just from $s\leftrightarrow -s$ symmetry). We conclude that our inclusion of the $\phi(\theta)\sim \theta^p$ configuration for $0<p<1$ is justified in the Schwarzian theory that comes from JT.

\subsubsection*{The case $p>1$}

Now we turn to the case $p>1$. Immediately from \eqref{eq:embeddiff} we see that this case will be more subtle: the fact that $\phi'\to 0$ as $\theta\to 0$ implies that $\eta$ cannot possibly remain finite; it must go to zero as $\theta$ goes to zero (so we will not have a true embedded Cauchy surface; it will touch the asymptotic conformal boundary $\scri_+$ at this point). Nonetheless, we will still be happy if we get a configuration with unambiguous finite action.

In this case, we are looking for solutions of $f'(s) = \frac{p-1}{s} - \sqrt{e^{2f(s)}-1} $. Generic initial conditions will have a good solution when solved in the direction of positive $s$, which converges on our desired Schwarzian limit as $s\to\infty$. But when we solve in the direction of decreasing $s$, they either blow up to infinity or hit $f=0$ in finite time. In neither case do we end up with a sensible Cauchy surface.

However, by tuning the initial conditions between these two possibilities we can find a unique solution which exists for all $s$. By making a logarithmic ansatz for $f$ as $s\to 0$, we find a unique approximate solution in a  small $s$ expansion:
\begin{equation}
    f(s) \sim \log\left(\frac{p}{s}\right) + \frac{s^2}{2p(p+2)}+\cdots
\end{equation}
where there are systematic corrections in even powers of $s$. This is verified by matching to a numerical solution.

Translating to $\eta(\theta)$, we find the behaviour at small $\theta\ll a^{-1}$ given by
\begin{equation}
    \eta(\theta) \sim -\theta^p(1+O(\theta^2)).
\end{equation}
So we indeed see that $\eta(0)=0$. But the extrinsic curvature has a similarly mild singularity $\mathcal{K}\sim \frac{p+1}{s}$ which leads to a finite Lagrangian and unambiguous finite action.

So, we conclude that the configuration of the dS JT path integral implied by the singular diffeomorphism $\phi(\theta)\sim\theta^p$ should be admissible: restoring the cutoff in JT regulates the divergent and ambiguous action of the Schwarzian theory. This justifies our choice of boundary conditions for the one-loop calculations in section \ref{sec:solution}.

\subsection{The case $p=1$ and the alternative boundary condition}

We would also like to check that the singular configurations of the alternative boundary condition (allowing $\phi(\theta) = A_+\theta$ for $\theta>0$ and $A_-\theta$ for $\theta<0$ with $A_+\neq A_-$) are ruled out in the context of  JT gravity. So, we repeat the analysis of off-shell configurations with $\phi(\theta) = A \theta$ (for $\theta>0$).

In this case it is simple to immediately solve \eqref{eq:embeddiff}:
\begin{equation}
    \eta(\theta) = - \frac{A}{a}\sin(a(\theta-\theta_0)).
\end{equation}
But this solution only makes sense in very small intervals $\theta_0<\theta<\theta_0+a^{-1}\pi$, and does not match onto the Schwarzian solution. Fortunately, there is one more extremely simple solution which achieves this, namely the constant $\eta = -\frac{A}{a}$.

However, if $A$ jumps discontinuously between $A_-$ and $A_+$, then the corresponding surface $\Sigma$ is also compelled to similarly jump: $\eta$ is discontinuous at $\theta=0$. This is clearly not a sensible Cauchy surface! Roughly, the singularity $b(\theta) \sim \delta'(\theta)$ that we encountered in section \ref{ssec:altbcs} corresponds to a $\delta'(\theta)$ singularity in extrinsic curvature $\mathcal{K}$, which is just a jump in the location of the surface. Importantly, this sort of singularity does not get regularised when we restore the cutoff $a$. We get similar jumps for $p\neq 1$ if we allow the coefficient of $|\theta|^p$ to be different for $\theta<0$ and for $\theta>0$. So, we see that the alternative boundary conditions (and in particular, the associated additional zero modes) are naturally ruled out for the Schwarzian theory coming from embedding into JT.

\section{Discussion}

We close with some open questions and directions for future work.

\subsection{A microscopic dual?}

The original $\Diff/PSL(2,\RR)$ Schwarzian theory has a microscopic realisation in the SYK model \cite{Maldacena:2016hyu,Sachdev:1992fk}, in which the Schwarzian arises as the effective action for a pseudo-Goldstone mode governing low-energy reparametrisations. What we have found in this paper immediately suggests several generalisations.

First, is there a similar microscopic realisation for all the other orbits (including the novel $T_{n,\Delta}$ and  $\tilde{T}_{n,\pm}$ theories? Second, our perspective has highlighted the need to consider varying couplings $u(\theta)$ (which may change sign). How do we describe this in SYK or another microscopic completion? One guess would be to simply consider a time-dependent Hamiltonian of the form $H(t) = u(t)H_{SYK}$, but this would not correctly account for the anomalous transformation law \eqref{eq:anomIntro} so this is too na\"ive. Additionally, we suspect that SYK should be governed by a different completion of the $SL(2,\RR)$ Schwarzian theory in the case that $u$ does not have a definite sign, as discussed in section \ref{ssec:discbcs}.

The varying coupling gives us novel results even for the original $\Diff/PSL(2,\RR)$ Schwarzian theories, in particular the striking $\delta(\int\frac{1}{u})$ dependence in \eqref{eq:SL2result} (assuming that the microscopic theory shares the same completion as JT gravity). How can we see these features in a microscopic model, which may serve as a candidate holographic dual for nearly $dS_2$ quantum gravity?

\subsection{Correlation functions}

In this paper we have only evaluated the Schwarzian path integral as a functional of the coupling $u$, but have not discussed  correlation functions with additional insertions in the path integral.

In de Sitter JT gravity, these are relevant for computing cosmological correlation functions \cite{Maldacena:2019cbz}. Note that in this case, the path integral with operator insertions computes the Wheeler-DeWitt wavefunction including matter fields (at least in perturbation theory). Computing correlation functions of operators defined at $\scri_+$ requires two copies of this calculation (for the `bra' and `ket' wavefunctions), glued together with an integral over couplings. In \cite{Maldacena:2019cbz}, this was simply an integral over constant couplings, corresponding to gauge-fixing on slices of constant dilaton $\Phi$ (and integrating over the volume of the slice), with an associated Klein-Gordon inner product (see \cite{Held:2025mai} for more on the relation between Klein-Gordon inner products and gauge fixing). But clearly this gauge is not applicable in the sectors where the renormalised dilaton changes sign: how do we modify this calculation with an appropriate gauge to apply to the other sectors?

The new results for the $SL(2,\RR)$ Schwarzian theory raise a simpler and apparently natural question about the Hartle-Hawking state of dS JT gravity. Namely, what is the relative probability of the two possible sectors (where the renormalised dilaton on $\scri_+$ is either everywhere positive, or changes sign)? Embedding in the near-Nariai $dS_2\times S^2$ context, this asks for the probability of forming a black hole singularity somewhere (rather than an everywhere-expanding universe). Admittedly, this question need not have a meaningful answer since the Hartle-Hawking state is non-normalisable in JT due to the dilaton zero modes. But this nonetheless remains an interesting first step to pursue (perhaps in the context of a modified dilaton potential\footnote{We would still have a Schwarzian theory at $\scri_+$ in a theory of dilaton gravity with asymptotically linear dilaton potential, though the potential would affect how a particular state (e.g., Hartle-Hawking) is described as a superposition over different constant curvature geometries near $\scri_+$, and hence over different orbits $\orbit$. But for the novel sectors we would require this linear potential at both $\Phi\to +\infty$ and $\Phi\to -\infty$ (and with the same slope, at least if we wanted to keep the same de Sitter radius in regions of $u>0$ and $u<0$). We would also need to  check whether the full shape of the potential leaves any mark (e.g., deforming our boundary conditions) at the zeroes of $u(\theta)$ where the dilaton runs through its full range.} to mitigate the divergent norm), towards relating the Schwarzian theory to cosmological observables in the new sectors.



\subsection{Alternative boundary conditions}\label{ssec:discbcs}

As discussed in section \ref{ssec:hyponeloop}, the failure of essential self-adjointness at zeroes of $u$ forces a choice of boundary conditions which is not uniquely determined by the formal path integral. We concentrated on  one natural choice guided by dS JT gravity, but also commented on another consistent option which leads to qualitatively different physics in section \ref{ssec:altbcs}. In that case, $2n-1$ zero modes enforce constraints on the coupling $u$, and the path integral receives contributions only on a restricted submanifold in coupling space. It would be interesting to understand whether this alternative has an independent physical realisation.

Furthermore, these two options do not exhaust all possible completions. There is one choice which is particularly natural from the perspective of SYK or AdS JT gravity (governed by the usual $SL(2,\RR)$ Schwarzian theory), in which context $\theta$ is interpreted as a time coordinate and the path integral computes the trace of some time-evolution operator.\footnote{I'd like to thank Douglas Stanford for bringing up this perspective, which led to the following observations.}  In the case of constant $u$ this is simply $\Tr(e^{-iHt})$ with $t \propto u^{-1}$, and as long as $u$ is never zero we can reduce to this case by passing to a `proper time' coordinate $t$ with $dt \propto \frac{d\theta}{u(\theta)}$. For real $u$ this `infinite temperature' trace is not well-defined (except in a distributional sense). Fortunately, in this case there is an obvious natural regulator using an $i\epsilon$ prescription, by deforming the coupling to $u(\theta) + i\epsilon$. This adds a small imaginary time or inverse temperature $\beta \propto \int \frac{\epsilon}{\epsilon^2+u^2}$ which we can take to zero after computing the path integral. In the case where $u$ is never zero, $\beta\to 0$ when $\epsilon\to 0$ and this reproduces the usual result given in the second line of table \ref{tab:results} (with a particular phase).

What if we allow $u$ to change sign (which we can interpret as a `timefold' calculation with branches of both forward and backward evolution), but continue to use this $i\epsilon$ prescription? The resulting Schwarzian path integral does not match the last line of table \ref{tab:results} (with the delta-function of $\int\frac{1}{u}$): instead, it simply gives the same result as the second line of the table, except that the integrals must all be evaluated using the $i\epsilon$ prescription. In this case, $\beta$ does not go to zero as we take $\epsilon\to 0$ with fixed smooth real $u$: instead, $\frac{1}{u(\theta)+i\epsilon}$ has imaginary delta-function contributions at zeroes $\theta_k$ of $u$, giving $\beta$ proportional to the sum of $|u'(\theta_k)|^{-1}$.\footnote{For a simple timefold where we take $u$ to be a step function which is locally constant but jumps between positive and negative values, this inverse temperature goes to zero. To see this, we smooth out the jumps over a narrow interval and note that $|u'(\theta_k)|\to\infty$ as we take the limit as the interval shrinks to zero size. We ultimately get a finite real result  for $\int\frac{1}{u+i\epsilon}$ which is independent of the way we choose to do the smoothing.} From the perspective of the one-loop calculation this prescription corresponds to a different boundary condition, defining an extension of $Q$ which is \emph{not} self-adjoint. In terms of the Frobenius solutions \label{eq:domQ}, the coefficient $c_1$ of the linear term jumps by an imaginary multiple of the coefficient $c_1'$ of the $\log$ term (as we get from defining the branch of $\log$ using the $i\epsilon$ prescription).

The conclusion is that the $SL(2,\RR)$ Schwarzian theory which computes the JT wavefunction in dS$_2$ (and in particular which gives the result in the final line of table \ref{tab:results}) is not the same as the theory which appears in SYK (or real-time calculations in AdS JT gravity) in this respect.

\paragraph{Acknowledgements}

I would like to thank  Kristan Jensen, Mukund Rangamani, Steve Shenker and Douglas Stanford for helpful discussions, comments and feedback on on a draft. I am  supported by DOE grant DE-SC0021085 and a Bloch fellowship from Q-FARM.

\appendix

\section{A direct approach to the one-loop determinant}\label{app:alt1loop}

We saw in section \ref{ssec:1loopsetup} that the one-loop calculation is na\"ively determined by a ratio of determinants $\frac{\det'{Q}}{\det'{\Omega}}$ where the primes denote exclusion of the unphysical zero mode $v\propto u$. Furthermore, we can write $Q = \Omega A$ with  $A v= uv' - u'v$, which suggests that we can simply cancel the $\Omega$ dependence and compute $\det'(A)$.

This approach is perfectly satisfactory and justifiable when $u$ has no zeroes. The eigenvalues $\lambda$ of $A$ are given by solutions to $Av=\lambda v$, which is explicitly solved by
\begin{equation}
    v(\theta) = u(\theta) \exp\left(\lambda \int^\theta\frac{1}{u} \right).
\end{equation}
For periodicity we require that $\lambda = \frac{2\pi i}{\int\frac{1}{u}} n$ for integer $n$. So these are the eigenvalues, with $n=0$ corresponding to the zero mode $v=u$ which we exclude (furthermore, the spectrum is identical to the set of eigenvalues for this operator: $A$ has purely point spectrum). Taking the product of these eigenvalues (using zeta function regularisation, for example) recovers the result
\begin{equation}
    \frac{\det'(Q)}{\det'(\Omega)} = \det{}'(A) \propto \int\frac{1}{u}.
\end{equation}
A similar approach works for the $SL(2)$ theories when $u(\theta)$ has no zeroes. The only change is that the $n=\pm 1$ modes are excluded as zero modes (along with $n=0$) so the result is $\left(\int \frac{1}{u}\right)^3$ as used in table \ref{tab:results}.

However, when $u$ has zeroes this becomes more subtle. The spectrum of $A$ (defined on the domain of smooth functions, or on the Sobolev space $H^1(S^1)$) is in fact the entire complex plane, so we do not recover a sensible determinant. We can make an ad-hoc prescription that restricts to eigenfunctions of the above form with a principle value prescription to define the integrals. But it not clear to us how to justify such a prescription in a principled way. This is why we followed the more laborious (but more principled) approach of the main text.

\section{Self-adjoint extensions}\label{app:ESA}

\subsection{Review of theory}

The central object in one-loop path integrals (and perturbation theory) is a quadratic form --- the second-order expansion of the action around a classical solution --- on an infinite-dimensional space of functions. To make sense of these integrals it is convenient to introduce an inner product to turn our function space into a Hilbert space: we will take the example of scalar functions on a compact manifold $M$, so our Hilbert space is $L^2(M)$ with some choice of measure. With this, we can write the quadratic action in terms of an operator $S$ on the Hilbert space:
\begin{equation}
    I_\mathrm{quad}[v] = \frac{1}{2}\langle v|Sv\rangle\,,
\end{equation}
where $S$ is typically some finite-order differential operator. Such $S$ will not be defined for all square-integrable functions, but only functions which are sufficiently smooth, and perhaps with additional conditions if the coefficients in $S$ have singularities. So $S$ is initially defined as $S:D(S)\to L^2(M)$ where the domain $D(S)\subset L^2(M)$ is dense, for example $D(S) = C^\infty(M)$ (or $D(S) = C^k(M)$ or a Sobolev space $D(S)=H^k(M)$ where $k$ is the order of $S$). On this domain, $S$ will be symmetric (or antisymmetric for the fermionic path integrals, in which case we can consider $iS$), meaning
\begin{equation}
     \text{$S$ symmetric} \iff \langle v'|Sv\rangle = \langle Sv'|v\rangle \qquad \forall v,v'\in D(S).
\end{equation}
However, this is not a sufficiently strong condition for us to unambiguously define determinants and inverses. We would like $S$ to have the desirable properties of a Hermitian matrix, namely a complete basis of eigenvectors with real eigenvalues. This is guaranteed (by the spectral theorem) if $S$ is self-adjoint $S^\dag = S$, or at least \emph{essentially self-adjoint}; we review these definitions next.

The adjoint $S^\dag$ is defined by the relation $\langle S^\dag f|v\rangle=\langle f|Sv\rangle$ whenever it makes sense, so we have
\begin{equation}
    f\in D(S^\dag) \text{ and }S^\dag f = g \iff \langle g|v\rangle = \langle f|Sv\rangle\quad \forall v\in D(S).
\end{equation}
This uniquely defines $S^\dag f$ if it exists (since $D(S)$ is dense). In the case of differential operators on $L^2(M)$ with $D(S)=C^\infty(M)$, this definition means that $S^\dag$ is given by weak (distributional) derivatives, in the case where this yields a square integrable function. Since $S$ is symmetric, we will certainly have $D(S)\subseteq D(S^\dag)$ and $S^\dag v = S v$ for $v\in D(S)$. But $S$ can fail to be self-adjoint because $D(S^\dag)$ can be strictly larger than $D(S)$. To resolve this we need to define $S$ on a bigger space, finding a \emph{self-adjoint extension} $\tilde{S}$ of $S$. An extension $\tilde{S}$ of $S$ is an operator with a larger domain $D(S)\subset D(\tilde{S})$, which agrees with $S$ on the original common domain ($\tilde{S}|v\rangle = S|v\rangle$ for $v\in D(S)$).

In many cases this distinction is trivial. If the adjoint $S^\dag$ is symmetric (so $\langle S^\dag v'|v\rangle=\langle v'|S^\dag v\rangle$ for all $v,v'\in D(S^\dag)$), then $S^\dag$ itself is a self-adjoint extension of $S$, and in fact is the unique self-adjoint extension. In this case, we say that $S$ is \emph{essentially self-adjoint}. But this does not always hold: there can be many incompatible self-adjoint extensions (or no self-adjoint extension at all, though this possibility will not be relevant for us). In practice, this  can happen because the proof of symmetry of $S$ uses integration by parts, and boundary terms which vanish for smooth $v$ may become non-zero when $v$ is less smooth.

To diagnose this, one method is to directly analyse the failure of the spectrum to be real. This means looking for eigenvectors $v$ with $S^\dag v = \lambda v$ for complex $\lambda\in\CC -\RR$. We can define the \emph{deficiency indices} $d_\pm$ by the dimension of the space of such eigenvectors:
\begin{equation}
    d_\pm(S) = \dim \ker (S^\dag - \lambda), \quad \operatorname{Im}(\lambda)\gtrless 0.
\end{equation}
These are independent of the choice of $\lambda$ in the upper/lower half-plane, so we can choose $\lambda = \pm i$ for example. The key result is that $S$ has a self-adjoint extension if and only if $d_+ = d_-=d$, and if this holds then the space of self-adjoint extensions is classified by elements of $U(d)$ (unitary maps $\ker(S^\dag - i)\to\ker(S^\dag + i) $). In our case, the operator $S$ will always be real (or pure imaginary for the fermionic measure integral), which already implies that $d_+=d_-$ so there will always be some self-adjoint extension.

\subsection{Practical strategy}

Now we explain how to turn this theory into a practical strategy, in our case where $S$ is a differential operator on the circle which we can write as
\begin{equation}
    S v(\theta) = \sum_{i=0}^k a_i(\theta) v^{(i)}(\theta)
\end{equation}
for some smooth functions $a_i$ (except perhaps at some isolated singularities). To evaluate the deficiency indices, we look for weak solutions to $Sv = \lambda v$ for general complex $\lambda$ (also requiring that $v$ is square integrable).

Now, weak solutions are smooth functions except at singular points where the coefficient of the leading term $a_k$ vanishes, or at singularities of any $a_i$. So we start by looking for smooth solutions in the neighbourhood of a singular point (which we can place at $\theta=0$ WLOG). In all cases that we encounter, these will be regular singular points which means that solutions can be constructed (locally) by the Frobenius method. This means making an ansatz $v(\theta)\sim \theta^\sigma(1+\cdots)$, where the $\cdots$ are a power series in $\theta$. Inserting this ansatz, the leading term as $\theta\to 0$ gives a polynomial equation for $\sigma$  (the indicial equation), which in all cases for us will be independent of $\lambda$. Solutions to this tell us how the independent solutions behave as $\theta\to 0$ (with an additional solution $v\sim \theta^\sigma \log\theta$ when the indicial equation has a repeated root). We exclude solutions with $\sigma \leq -\frac{1}{2}$, because they fail to be square integrable.

So, a general $L^2$ weak solution will be a linear combination of the $k$ Frobenius solutions (excluding non-normalisable solutions) in the region $\theta>0$ to the right of the singular point, and also in the region $\theta<0$ to the left. But this is not sufficient; we will typically have additional conditions relating the solutions on either side (otherwise, $Sv$ may contain derivatives of delta functions). To check this carefully, we evaluate $\int v (S-\lambda) f$ for a general smooth function $f$ (perhaps with additional conditions on its derivatives to guarantee that $Sf$ is smooth, if some of the $a_i$ are singular). Because $S$ is symmetric this can be evaluated by integration by parts, at the cost of boundary terms at the singular points. In more detail, the integral can also be written as the limit of integrals $\int_\epsilon v (S-\lambda) f$ where we exclude some interval of size $\epsilon>0$ containing the singular points, and take $\epsilon\to 0$ at the end; symmetry of $S$ implies that $v(S-\lambda)f = vSf-fSv$ is a total derivative, so our integral can be expressed as differences of limits as $\theta\to 0$ from the left and right (along with contributions from other singular points). These will be proportional to $f(0)$, $f'(0)$ and so forth; for a weak solution, the coefficients of these must all vanish.

This will give us some number of conditions matching the coefficients of the Frobenius solutions to the left and right of the singular point. If the number of conditions is smaller than the order $k$ of the equation, then the equation is underdetermined, and the difference is the number of additional boundary conditions we must choose. The deficiency index is just the sum of this number over all singular points.

\subsection{Examples}

Now we implement the above strategy for each of the relevant examples in the main text.

\subsubsection{$Q$ at a simple zero of $u$}
First, we look at an operator
\begin{equation}
    Q v = \theta v''''(\theta) + 2 v'''(\theta)+\cdots,
\end{equation}
which appears in the main text when $u$ has a simple zero. The $\cdots$ terms are unimportant for this analysis (as long as they are nonsingular), so we disregard them. The indicial polynomial is $\sigma(\sigma-1)^2(\sigma-2)$, so the general solution to $S$ goes as
\begin{equation}
    v(\theta)\sim c_0 + c_1\theta + c_1' \theta \log|\theta| + c_2 \theta^2 + \cdots
\end{equation}
as $\theta\to 0$. Symmetry of $S$ follows from the identity
\begin{equation}
    vS f - f S v \sim (\theta  f''' v-\theta  f v'''+\theta  f' v''-\theta  f''v'+v f''-f v'')',
\end{equation}
and the $\theta\to 0$ limits of the total derivative on the right will give us our boundary conditions on the coefficients. Taking $f$ to be compactly supported in a region containing $\theta=0$ (so other singular points do not contribute), we find
\begin{equation}
    \int v (S-\lambda) f = \Delta c_0 f''(0) + \Delta c_1' f'(0) - 2\Delta c_2 f(0),
\end{equation}
where $\Delta$ denotes the difference in the given coefficient between $\theta>0$ and $\theta<0$. So, we find the three conditions that $c_0$ $c_1'$ and $c_2$ match on either side, but no condition on $c_1$. This singular point contributes $1$ to the deficiency index; we should choose one extra boundary condition here to define a self-adjoint extension. Our choice is the obvious $\Delta c_1 = 0$, and we also consider the alternative $c_1'=0$.

\subsubsection{$Q$ at a double zero of $u$}
For the one-loop expansion around smooth solutions for the parabolic orbits, we consider a double zero of $u$ (with $u'''=0$) giving an operator of the form
\begin{equation}
    Q v = \theta^2 v''''(\theta) + 4\theta v'''(\theta)+\cdots.
\end{equation}
The indicial equation is $(\sigma+1)\sigma(\sigma-1)(\sigma-2)$. The $\sigma=-1$ solution isn't square integrable, so must be excluded. The boundary conditions tell us that $v$ and $v'$ are continuous, but $v''$ can jump: we can have singularities of the form $\theta^2 \Theta(\theta)$. This operator is essentially self-adjoint.

 Locally, zero modes are $u(\theta) p(\int^\theta\frac{1}{u})$ where $p$ is a cubic polynomial. To avoid non-normalisable singularities the cubic term must be absent, so $p$ is in fact quadratic. The coefficients of the quadratic and linear terms must be constant as we cross the singularity (with a principle value definition of the integral). If $\int\frac{1}{u}\neq 0$, these terms must be absent for a periodic solution. But, the constant term can jump. This operator gives rise to the version of the parabolic theories discussed in section \ref{ssec:altbcs}.

 \subsubsection{$\Omega,Q$ with singular $b$}

Now if consider what happens near a singularity where $b\sim \frac{c}{24}\frac{p^2-1}{\theta^2}$. This includes the case when 

In this case we have a singular point even for the measure operator $\Omega$:
 \begin{equation}
     \Omega v = v''' +\frac{1-p^2}{\theta^2}v'-\frac{1-p^2}{\theta^3}v
 \end{equation}
 with $p>0$ and $p\neq 1$. Because of the singularities, this is not a well-defined operator acting on every smooth function: the original domain $D(\Omega)$ consists only of smooth functions with $f(0)=f''(0)=0$. The indicial equation gives Frobenius exponents $\sigma = 1,1+p,1-p$. For $p<\frac{3}{2}$ these are all square integrable; for $p\geq \frac{3}{2}$ we must exclude the $\theta^{1-p}$ solution. The boundary conditions imply only that the coefficient of the smooth $\theta$ solution is continuous. So, $\Omega$ is not essentially self-adjoint for $p<\frac{3}{2}$: we must impose two extra boundary conditions at the singular point. The obvious condition defining our self-adjoint extension is simply equality between coefficients of $|\theta|^{1 - p}$ and $|\theta|^{1 + p}$ in the $\theta\to 0$ limit from left and right, and this is justified from the action of singular diffs \eqref{eq:singdiff}.
 
 For $p\geq \frac{3}{2}$, normalisability would exclude the desired boundary condition since $|\theta|^{1-p}$ becomes non-normalisable. So in this case our general strategy of defining $\Omega,Q$ as self-adjoint operators on $L^2(S^1)$ cannot result in the desired definition of the theory (e.g., transforming as \eqref{eq:anom} under our non-smooth diffs and remaining invariant under the deformation used for localisation in section \ref{sec:localisation}).

 \subsubsection{$Q$ with singular $b$}

Now we repeat the above with $Q$, defined by acting with $\Omega$ on $A v(\theta) = \theta v'(\theta) - v(\theta)$. The original domain once again can be taken to consist of smooth $f$ such that $f(0)=f''(0)=0$. The indicial polynomial is $(\sigma-1)^2(\sigma-1-p)(\sigma-1+p)$, giving solutions going as $\theta$, $\theta\log|\theta|$ and $|\theta|^{1\pm p}$. Demainding a weak solution only implies a single boundary condition: continuity of the coefficient of $\theta\log|\theta|$. Finding a self-adjoint extension here is a hybrid of the exercise for $\Omega$ immediately above, and for the case of $Q$ at a simple zero but without a singularity.

\bibliographystyle{JHEP-2}
\bibliography{SchwarzBib}

\end{document}